\documentclass[]{aa}  

\usepackage{txfonts}
\usepackage{newtxtext,newtxmath}
\usepackage[T1]{fontenc}
\usepackage{graphicx}   
\usepackage{amsmath}    
\usepackage[normalem]{ulem}
\usepackage{upgreek}
\usepackage[shortlabels]{enumitem}
\usepackage{float}
\usepackage{xcolor}
\usepackage{natbib}
\usepackage[linkcolor=blue]{hyperref}
\usepackage{subcaption}
\usepackage{soul}
\usepackage{array}
\usepackage{multirow}
\usepackage{soul}
\usepackage[switch]{lineno}

\newcommand{\simname}[1]{\texttt{#1}}
\newcommand{\bl}[1]{\mbox{\boldmath$ #1 $}}

\let\oldpageref\pageref
\renewcommand{\pageref}{\oldpageref*}

\begin{document}

   \title{Magnetic disk winds in protoplanetary disks:}

   \subtitle{Description of the model and impact on global disk evolution}

   \titlerunning{Winds in protoplanetary disks I}

   \author{Kundan Kadam\inst{1,3} \thanks{email: kundan.kadam@oeaw.ac.at},
          Eduard Vorobyov\inst{2},
          Peter Woitke\inst{1},
          Shantanu Basu\inst{3,4},
          \and
          Sierk van Terwisga\inst{1}
          }

   \institute{Space Research Institute, Austrian Academy of Sciences, Schmiedlstrasse 6, A-8042 Graz, Austria
   \and
   Department of Astrophysics, The University of Vienna, A-1180 Vienna, Austria
   \and
    Department of Physics and Astronomy, University of Western Ontario, London, Ontario, N6A 3K7, Canada
    \and
    Institute for Earth \& Space Exploration, University of Western Ontario, London, Ontario, N6A 5B7, Canada
             }

   \date{Received September 15, 1996; accepted March 16, 1997}

  \authorrunning{Kadam K. et. al}
  
  \abstract
   {
   Canonically, a protoplanetary disk is thought to undergo (gravito-)viscous evolution wherein the angular momentum of the accreting material is transported outward. 
   However, several lines of reasoning suggest that the turbulent viscosity in a typical protoplanetary disk is insufficient to drive the observed accretion rates.
   An emerging paradigm suggests that radially extended magnetic disk winds, which transport angular momentum vertically, may play a crucial role in disk evolution.
   }
   {We propose a global model of magnetic wind-driven accretion for the evolution of protoplanetary disks in the thin-disk limit based on the insights gained from local shearing box simulations. 
   In this paper, we aim to develop this model and constrain the model parameters with the help of theoretical expectations and through comparison with observations.}
   {
   The magnetic wind is characterized with the associated loss of angular momentum and mass, and we modeled these with fitting formulae that depend on the local disk conditions and stellar properties. 
   We incorporated the disk winds self-consistently in the numerical magnetohydrodynamic code FEOSAD and studied the formation and long-term evolution of protoplanetary disks. 
   We included disk self-gravity and an adaptive turbulent $\alpha$ that depends on the local ionization balance, while the co-evolution of a two-part dusty component was also considered.
   We obtained synthetic observations via detailed modeling with the radiation thermo-chemical code {\sc ProDiMo}.
   }
   {The models that include disk winds satisfy the general expectations from both theory and observations.
   The disk wind parameters can be guided by observational constraints, and the synthetic observations resulting from such a model compare favorably with the selected ALMA survey data of Class II disks. 
   The proposed magnetic disk wind model is a significant step forward in the direction of representing a more complete disk evolution, wherein the disk experiences concurrent torques from viscous, gravitational, and magnetic wind processes.}
   {}


   \maketitle
%

\section{Introduction}
\label{sec:intro}

During its formation, a low mass star accretes through a flattened centrifugally supported circumstellar disk. 
Such a disk inevitably forms due to the angular momentum contained in the parent molecular cloud or through local subsonic turbulence \citep{Yorke+93,PP6-Li,Verliat+20}. 
For the accretion of gas to proceed through the disk, the accreting material must efficiently lose its angular momentum.
{Gravitational instability (GI) facilitates angular momentum redistribution during protostellar disk formation from a magnetized prestellar core \citep{Tsukamoto+15,Masson+16,Xu-Kunz21,Mauxion+24} and continues to play a dominant role during the early embedded phases of the disk's evolution \citep{VB06,VB09, Kratter-Lodato16,Rice16}. 
This occurs when the disk is sufficiently massive as compared to the host star, and the self-gravity of the disk leads to the formation of spiral arms, density fluctuations, and fragmentation, resulting in efficient global mass transport. }

Later in the Class II stage, a protoplanetary disk (PPD) is canonically thought to evolve viscously, with magnetorotational instability (MRI) providing the turbulence necessary for this viscosity \citep{SS73,Pringle81,Balbus-Hawley91,Turner+14}. 
When a weak magnetic field embedded in the disk interacts with the partially ionized material, the Keplerian shear in the disk gives rise to MRI.
The action of MRI requires sufficiently ionized gas, and cosmic rays are thought to be the primary source of ionization in PPDs \citep{UN81, UN88,Thi+19}.
The cosmic rays penetrate the disk surface from the outside, passing through the entire disk thickness at large radii.
The innermost regions have a high enough surface density that the cosmic rays cannot reach the midplane, resulting in insufficient ionization to couple to the magnetic field.
This suppresses the MRI and causes the PPD to accrete through a layered structure, wherein the accretion occurs only through the ionized surface layers, and a magnetically dead zone is formed at the midplane \citep{Gammie96}.
The dead zone forms a bottleneck for mass and angular momentum transport and gives rise to such disk substructures as rings and vortices, which form pressure maxima and sites for dust growth \citep{Dzyurkevich10,Pinilla+12,Kadam+22,Regaly+23}.
Spatially resolved images of PPDs indeed show an abundance of disk substructure in the {majority of large, bright disks} \citep{Andrews+09,Flock+15,Andrews+18,Zhang+18,MAPS21,Hsieh+24}.

However, several lines of reasoning point toward a dramatically different picture of PPD evolution, at least in the Class II stage. 
The Shakura-Sunyaev $\alpha$-parameter quantifies the efficiency of angular momentum transport in the standard theory of viscous accretion \citep{SS73}.
In the absence of a dead zone, accretion within the disk is most efficient, and a typical value of $\alpha$ is about 0.01 \citep{Hartmann+98,Hughes+11,Andrews+10}.
However, recent observations of protoplanetary disks indicate that the inferred value of the $\alpha$ parameter is much lower, which is insufficient for the disk to accrete within its typical observed lifespan of a few million years \citep{Williams-Cieza11}.
For example, the dust rings observed within PPDs by the Atacama Large Millimeter/ Submillimeter Array (ALMA) are significantly thin, and studies of dust properties suggest the turbulence to be $\alpha \sim 10^{-3}-10^{-4}$ \citep{Pinte+16,Dullemond+18,Muro-Aren+18,Franceschi+23}.
However, some uncertainty still exists from the assumption of dust properties since the dusty rings can only constrain the ratio of $\alpha$ to the Stokes number.
Spatially and spectrally resolved molecular line measurements also indicate that the contribution of turbulence to the line broadening is consistent with weak turbulence \citep{Flaherty+15,Flaherty+17,Flaherty+20}.
Conversely, from a theoretical viewpoint, the disks in recent magnetohydrodynamic (MHD) simulations also point toward marginal excitation of MRI turbulence.
When the non-ideal MHD effects are included, especially ambipolar diffusion, the simulations show an increasingly laminar flow and much less overall turbulence. 
Instead, magnetic disk winds are launched near the disk surface, which carry a significant amount of angular momentum \citep{Suzuki-Inutsuka09,BS13,Fromang+13,Simon+13,Bethune+17,Zhu-Stone18,Cui-Bai20,Lesur21b}.

The mechanism for angular momentum loss via disk winds was first described by \cite{BP82}, and it is similar to magnetic braking of rapidly rotating stars. 
When the disk magnetic field makes a sufficiently large angle with the normal ($> 30^\circ$), the outflowing gas, which is constrained to move along the open magnetic field lines, is accelerated centrifugally.
The gas flow eventually bends the poloidal magnetic field component into a toroidal field, and the gradient of the toroidal magnetic pressure assists in the collimation of the wind \citep{Pudritz19}. 
We note that these magnetocentrifugal winds are fundamentally different from photoevaporative winds; the latter are launched thermally from the disk surface due to external irradiation and do not exert a torque on the disk \citep{Hollenbach+94,Alexander+14,Ercolano-Pascucci17}.
Photoevaporation is negligible in the early disk evolution, although it becomes significant in the late stages, when the stellar accretion rate becomes comparable to the photoevaporation rate.
Photoevaporative winds are considered to be responsible for the eventual dispersal of the gas in PPDs \citep{Alexander+06,Alexander+14}.
In reality, these two mass ejection processes operate simultaneously and can be collectively considered as a ``magneto-thermal wind" \citep{Bai+16,Lesur+22}.

High spectral resolution observations of gas lines that trace unbound disk material (forbidden lines such as [OI] $\lambda 6300$ and [SII] $\lambda 6731$) show the ubiquitous presence of outflows around PPDs \citep{Pascucci+23}.
These outflows are divided into a high-velocity component (HVC) traveling at hundreds of kilometers per second and a low velocity component (LVC) of 10-50 ${\rm km\,s^{-1}}$ \citep{Natta+14,McGinnis+18,Nisini+18,Rab+22,Fang+23}.
The HVC is interpreted as a collimated axial jet, which is launched very close to the surface of the accreting object via magnetic processes. 
The LVC component originating beyond the dust evaporation radius and spanning a majority of the disk is typically identified as magnetocentrifugal wind, although it may also be interpreted as X-ray driven photoevaporative flow \citep{Rab+23}.
Spatially resolved outflows of young objects show small-scale rotating flows as slow molecular winds in addition to fast moving jets \citep{Gudel+18,Lee+21}.
{These winds show a nested onion-like morphology at large opening semi-angles (20-50$^\circ$) and may originate within the inner ten au of the disk \citep{deValon+20,Pascucci+25}.}
The mass flux estimates of the disk winds are exceedingly significant, with one to ten times the jet mass flux and 0.3 to three times the accretion rate onto the central protostar \citep{Tabone+20}.
Thus, an emerging paradigm of protoplanetary disk evolution suggests that in addition to redistribution of angular momentum through gravitoviscous torques, extraction of mass and angular momentum via magnetic disk winds plays a crucial role as a driver of accretion.

As the winds fundamentally arise from the interaction of partially ionized gas with the magnetic field, theoretical insights into disk winds primarily come from 3D non-ideal MHD simulations.
This includes both local shearing box studies \citep{BS13,Simon+13,Suzuki-Inutsuka09} and global simulations, which resolve a significant fraction of the disk \citep{Bethune+17,Zhu-Stone18,Cui-Bai20,Lesur21b}. 
Since long integration times of the order of the disk lifetime (a few million years) are not feasible with 3D simulations, disk winds can be studied with simplified semi-analytical global evolutionary models in 1D \citep{Kunitomo+20,Alessi-Pudritz22,Somigliana+22,Tabone+22}. 
However, certain important processes occurring in PPDs cannot be faithfully represented in 1D axisymmetry.
For example, in the early stages of disk evolution, GI leads to the formation of large-scale spirals and clumps, which may eventually form wide orbit
gas giants \citep[][]{Boss97,VB10a,Kratter-Lodato16}.
In the later stages of evolution, disk-planet interactions are also intrinsically axially asymmetric \citep{Paardekooper+23}.
Thus, theoretical studies of long-term accretion disk evolution are often performed in the plane of the disk ($r,\phi$) in 2D, with the thin-disk approximation. 
The vertical hydrostatic equilibrium is imposed throughout the disk so that the vertical temperature gradients and motions are neglected. Such an approximation typically works very well for PPDs since the ratio of gas scale height to the local radius ($H_{\rm g}/r$) is much less than unity.
This approach is powerful for studying the long-term behavior of PPDs because it relaxes many of the simplifying assumptions of the 1D \cite{Pringle81}-like viscous equation of disk evolution while also being computationally feasible.

Almost all of the thin-disk models of protoplanetary disks to date assume a viscous evolution of the disk. The disk self-gravity is often considered, but magnetic winds are essentially neglected.
Recently, \cite{Kimmig+20} and \cite{Elbakyan+22} have considered the effects of wind-driven accretion in the thin-disk limit in the context of planetary migration and gap opening, respectively.  
Nevertheless, these models leave much room for improvement. 
The wind parameters, such as its escape velocity and the angle of the magnetic field, cannot be held constant over the disk lifetime. 
Similarly, the vertical angular momentum transport of magnetic disk winds is fundamentally different from the \cite{SS73} $\alpha$ parameter and cannot be modeled as an additional viscosity, as the latter only redistributes the angular momentum within the disk.
In addition, the co-evolution of the dust component is not explicitly considered.
Furthermore, accretion variability during star formation complicates the situation, especially the powerful FUor-type outbursts, wherein the accretion luminosity of the protostar increases by hundreds of times \citep{Hartmann-Kenyon96,Audard+14,Fischer+23}.
These events are important over the long term because they can significantly alter the disk composition, dynamics, and snow lines, while as much as 10-35\% of the main-sequence stellar mass can be accreted via episodic accretion \citep{Offner-McKee11,DV12,Fischer+19,Molyarova+18,Vorobyov+20,Vorobyov+22}.
The accretion variability during these events should also directly lead to a corresponding variability in the wind outflow rate \citep{Konigl-Pudritz00,Ellerbroek+14}. 
With these issues in mind, a model framework of magnetic disk winds for protoplanetary disks in the thin-disk limit, which incorporates all of the aforementioned phenomena, is desirable.

In this work, we propose a new global model of magnetic disk winds that takes into account the insights gained from local shearing box MHD simulations.  
The results of the shearing box simulations conducted at several radial distances can be stitched together in order to estimate the wind behavior throughout the PPD \citep{Bai13,BS13}.
We improve upon the wind mass loss rate and surface stress estimates of Bai et. al. by taking into account the corrections that are essential for long-term disk evolution.
The resulting prescription of winds is incorporated self-consistently into the global MHD code FEOSAD, which is capable of simulating the formation and evolution of PPDs over long timescales in the thin-disk limit \citep{VMHD20}.
Thus, all three major sources of angular momentum transport are now included, namely, turbulent viscosity, gravitational instability, and magnetic disk winds.
We follow the disk evolution from the earliest embedded Class 0 or I stages through to the later Class II stage, demonstrating that the proposed model of magnetic disk winds is a reasonable approximation in the thin-disk limit.
We confirm that the disk winds have a significant impact on the disk evolution and global disk properties. Most notably, the disks with winds tend to be smaller in both size and mass.
The simulation results are post-processed with the radiation thermo-chemical disk code {\sc ProDiMo} \citep{Woitke2009,Woitke2016}, and the obtained synthetic observations are compared with ALMA survey data of Class II disks.
Our analysis demonstrates that such a comparison can be used to guide and constrain the disk wind model parameters.

In the interest of the conciseness and readability of the study, we present in this paper our disk wind model in detail and explain the steps of obtaining synthetic observations and the subsequent comparison with large-scale surveys.
In a second companion paper, we will comprehensively investigate the properties of magnetic disk winds, for example, characteristic bulk and radial properties of the winds, their effects on substructure formation, and comparison of different modes of angular momentum transport.
The structure of the current paper is as follows. 
Section \ref{subsec:disk} describes our MHD disk evolution code, and Sect. \ref{subsec:wind} elaborates on the magnetic wind model.
In Sect. \ref{subsec:prodimo}, we explain how the synthetic observations are obtained, while Sect. \ref{subsec:init} describes the initial conditions.
In Sect. \ref{sec:results}, the results of the numerical simulations as well as radiation thermo-chemical modeling are presented in the context of observational constraints and theoretical consensus. The findings are summarized in Sect. \ref{sec:conclusions}.

\section{Methods}
\label{sec:methods}

\subsection{Disk evolution with FEOSAD}
\label{subsec:disk}

In this section we summarize the underlying MHD model that is used for the evolution of the PPD in the thin-disk limit.
The magnetic disk winds described in Sect. \ref{subsec:wind} are added to this base model in a modular fashion.
The hydrodynamic simulations are conducted using the code Formation and Evolution Of a Star And its circumstellar Disk \citep[FEOSAD,][]{VDust18,VMHD20}.
The numerical implementation of hydrodynamic algorithms is based on the operator-split procedure similar in methodology to the ZEUS code \citep{Stone-Norman92}.
We take into account the self-gravity of the disk, co-evolution of the coupled two-part dust component, calculation of magnetic fields in the flux-freezing approximation, and a dead zone modeled with an adaptive-$\alpha$ formulation, where the effective $\alpha$ is derived from ionization-recombination balance equations.  
Here we describe the minimum set of equations, while a more detailed description of the model can be obtained in \cite{VMHD20} and references therein.

The equations of mass continuity, conservation of momentum, and energy transport for the gas component in the thin disk limit can be described as follows:

\begin{equation}
\label{eq:cont}
\frac{\partial \Sigma_{\rm g}}{\partial t}  + \nabla  \cdot \left(\Sigma_{\rm g} \bl{v} \right)  = -\dot{\Sigma}_{\rm w,g}\, ,  
\end{equation}
\begin{eqnarray}
\label{eq:mom}
\frac{\partial}{\partial t} \left( \Sigma_{\rm g} \bl{v} \right) + \nabla \cdot \left( \Sigma_{\rm g} \bl{v} \otimes \bl{v} \right)  =   - \nabla {\cal P}  + \Sigma_{\rm g} \, \left( \bl{g} +\bl{g}_\ast \right) + \nabla \cdot \mathbf{\Pi} \nonumber \\
- \Sigma_{\rm d,gr} \bl{f}_{\rm b} +  \frac{B_z {\bl B}_p^+ }{2 \pi} - H_{\rm g}\, \nabla \left(\frac{B_z^2}{4 \pi}\right) -{ T}^{\rm Max},
\end{eqnarray}
\begin{equation}
\frac{\partial e}{\partial t} +\nabla \cdot \left( e \bl{v} \right) = -{\cal P}
(\nabla \cdot \bl{v}) -\Lambda +\Gamma + \nabla \bl{v}:{ \Pi} - \dot{e}_{\rm w}. 
\label{eq:energy}
\end{equation}
We note that the equations are solved in the plane of the disk in polar coordinates $(r,\phi)$ so that the vectors and tensors have only the planar components.
Here $\Sigma_{\rm g}$ is the gas surface density, $\bl{v}$ is the gas velocity in the disk plane and $\dot{\Sigma}_{\rm w,g}$ is the rate of mass loss (per unit disk area) from magnetic winds. 
The vertically integrated gas pressure, $\cal P$, is calculated via the ideal equation of state: ${\cal P}=(\gamma-1)e$, where $e$ is the internal energy per surface area and the assumed value of $\gamma$ is 7/5.
The term ${ g_\ast}$ is the gravitational acceleration by the central protostar, while the gravitational acceleration in the disk plane, $\bl{g}$, takes into account self-gravity of both the gaseous and the dusty components by solving the Poisson equation \citep[see][]{BT08}. 
Thus, the gravitational torques, which are a major source of mass and angular momentum transport in the early stages of disk evolution, are considered self-consistently.
The turbulent viscosity is taken into account via the viscous stress tensor ${ \Pi}$ \citep{VB10},  while the Maxwell's stress tensor, ${ T}^{\rm Max}$ originates in the disk winds.
The only component relevant for the surface stress is $T^{\rm Max}_{z\phi}$, which forms a sink term in the $\phi$ component of the momentum equation.
The term $\bl{f}_{\rm b}$ is the drag force per unit mass due to the back-reaction of dust on to the gas, and $\Sigma_{\rm d,gr}$ is the surface density of grown dust.
Later in this section, the same back-reaction term enters the dynamics of grown dust.
The vertical scale height of the gas disk, $H_{\rm g}\!=\!c_{\rm s}/\Omega$, is calculated assuming vertical hydrostatic balance. 
The magnetic field in the disk has only a vertical component $B_z$, while planar components exist at the top and bottom surfaces.
The planar component of the magnetic field at the top and bottom surfaces of the disk are denoted by ${\bl B}_p^+$ and ${\bl B}_p^-$, and a symmetry about the disk midplane is assumed, implying ${\bl B}_p^-=- {\bl B}_p^+$.
This approach of evolving the magnetic field is similar to evolution of a thin magnetized disk \citep{CM93,BM94,VB06}.

The energy transport equation (Eq. \eqref{eq:energy}) pertains to the evolution of the internal energy per surface area, $e$. 
The disk midplane temperature ($T_{\rm mp}$), which is assumed equal for the dust and gas components, is obtained by solving this energy balance equation.  
The heating ($\Gamma$) and cooling ($\Lambda$) rates are based on the analytical solution of the radiation transfer equations in the vertical direction \citep{Dong+16}.
The heating function takes into account the irradiation of the disk surface from the stellar as well as background irradiation and radiation cooling by dust grains is also considered \citep{VDust18}.
{ The irradiation from the central star is calculated using incidence angle of radiation arriving at the disk surface, which in turn is obtained using the local scale height and a fixed flaring angle \citep{VB10,Dong+16}.
These calculations neglect the self-shielding of the disk, which may arise in the innermost regions due to formation of concentric rings. 
However, this effect is considered to be insignificant, as viscous heating dominates the energy balance in the midplane of this inner disk region. 
}
The resulting model has a flared structure, wherein the disk vertical scale height increases with radial distance.
Both the disk and the infalling envelope receive a fraction of the irradiation energy from the central protostar, both in terms of accretion and photospheric luminosity \citep{VB09}.
The optical depths in the calculations are proportional to the total dust surface density; however, they do not take into account the dust growth.
The Planck and Rosseland mean opacities needed for these calculations are taken from \cite{Semenov+03}. 
The term involving viscous stress tensor accounts for the viscous heating, which typically dominates the inner disk and becomes significant for an extended region during outbursts \citep{VB10,Kadam+20}.
The last term in Eq. \eqref{eq:energy} corresponds to the internal energy of matter carried away by the magnetic winds.

The dust is modeled with two components and evolves via the processes of coagulation, fragmentation, and drift \citep{VDust18}. 
The small dust component is assumed to be fully coupled with the gas, while the grown dust drifts with respect to the gas and contributes to the back-reaction term in Eq. \eqref{eq:mom}. 
The small dust has a grain size of $a_{\rm min}<a<a_\ast$, and grown dust corresponds to a size of $a_\ast \le a<a_{\rm max}$, where $a_{\rm min}=5\times 10^{-3}$~$\upmu$m and $a_\ast=1$~$\upmu$m. 
Here, $a_{\rm max}$ is a dynamically varying maximum size of the dust grains, which depends on the efficiency of radial dust drift and the rate of dust growth.
The dust grains are assumed to have a constant density of $\rho_{{\rm s}}=3.0\,{\rm g~cm}^{-3}$.
The equations of continuity and momentum conservation for small and grown dust components are
\begin{equation}
\label{eq:contDsmall}
\frac{{\partial \Sigma_{\rm d,sm} }}{{\partial t}}  + \nabla  \cdot 
\left( \Sigma_{\rm d,sm} \bl{v} \right) = - S(a_{\rm max})  - \dot{\Sigma}_{\rm w, d},  
\end{equation}
\begin{equation}
\label{eq:contDlarge}
\frac{{\partial \Sigma_{\rm d,gr} }}{{\partial t}}  + \nabla  \cdot 
\left( \Sigma_{\rm d,gr} \bl{u} \right) = S(a_{\rm max}),  
\end{equation}
\begin{eqnarray}
\label{eq:momDlarge}
\frac{\partial}{\partial t} \left( \Sigma_{\rm d,gr} \bl{u} \right) +  \nabla \cdot \left( \Sigma_{\rm
d,gr} \bl{u} \otimes \bl{u} \right)  &=&   \Sigma_{\rm d,gr} \, \left( \bl{g} + \bl{g}_\ast \right) + \nonumber \\
 + \Sigma_{\rm d,gr} \bl{f}_{\rm b} + S(a_{\rm max}) \bl{v},
\end{eqnarray}
where $\Sigma_{\rm d,sm}$ and $\Sigma_{\rm d,gr}$ are the surface densities of small and grown dust, respectively.
The term $\bl{u}$ denotes the planar components of the grown dust velocity and $\dot{\Sigma}_{\rm w, d}$ accounts for the small dust blown away with the disk winds.
Since the winds originate in the upper layers of the disk surface, we assume that they do not carry grown dust. 
$S$ is the rate of conversion from small to grown dust per unit surface area, which is a function of the maximum size of the dust $(a_{\rm max})$.
The dust is assumed to mix vertically with the gas, which is a reasonable approximation for a young disk evolving under gravitational and viscous torques.
$S(a_{\rm max})$ is derived from the assumption that the size distributions of both the dust populations can be described by a power law with an exponent of $-3.5$.  
The discontinuity in the dust size distribution at $a_*$ is assumed to get smoothed out, which implies the dominant role of dust growth as compared to the dust flow \citep[][]{Molyarova+21}.
The evolution of the maximum size of the grown dust can be expressed as an advection equation modified for the presence of a source term:
\begin{equation}
\frac{\partial a_{\rm max}} {\partial t} + ({\bl u} \cdot \nabla) a_{\rm max} = \cal{D}.
\label{eq:dustA}
\end{equation}
Here,
$\cal{D}=\rho_{\rm d} {\it v}_{\rm rel}/\rho_{\rm s}$
is the growth rate that accounts for coagulation and $\rho_{\rm d}$ is the total dust volume density.
{  The relative dust-to-dust velocity, $v_{\rm rel}$, is calculated by considering the main sources of relative velocities -- the Brownian motion and turbulence \citep[][]{VDust18}.}
The maximum size that the dust can achieve is limited by the fragmentation barrier
\begin{equation}
 a_{\rm frag}=\frac{2\Sigma_{\rm g}v_{\rm frag}^2}{3\pi\rho_{\rm s}\alpha_{\rm turb} c_{\rm s}^2},
 \label{afrag}
\end{equation}
where $v_{\rm frag}$ is the fragmentation velocity, $c_{\rm s}$ is the sound speed and { $\alpha_{\rm turb}$ corresponds to the turbulence parameter \citep{Birnstiel+12}, wherein the possible contribution from gravitoturbulence is not considered \citep[see, e.g.,][]{2023A&A...670A..81V}}. 
{ We set the fragmentation velocity to a canonical value of 3 ${\rm m\,s^{-1}}$, although there is significant ambiguity due to sticking properties of different grain compositions, for example, bare silicate versus icy grains \citep{Blum-Wurm08, Wada+09, Steinpilz+19}.}
{ When $a_{\rm max}$ exceeds $a_{\rm frag}$, the growth rate $\cal{D}$ is set to zero and $a_{\rm max}$ is set equal to $a_{\rm frag}$, which guarantees that the particle size is capped at $a_{\rm frag}$.}
We note that in our approach, the dust growth is limited to keep the size of dust particles within the Epstein regime so that the back-reaction term is expressed as
\begin{equation}
     \bl{f}_{\rm b} = \frac{\bl{v}-\bl{u} }{t_{\rm stop}}, 
      \label{eq:f_p}
\end{equation}
where the stopping time is $t_{\rm stop} = a_{\rm max} \rho_{\rm s}/ c_s \rho_{\rm g}$. 
The update of gas and dust velocities due to the friction force is calculated using a fully implicit scheme \citep{Stoyanovskaya+18,Stoyanovskaya+20}.

The viscosity in a protoplanetary disk is thought to originate primarily due to the turbulence generated by MRI \citep{Balbus-Hawley91,Turner+14}. 
As the cosmic rays externally penetrate the disk surface, the inner disk ($\lesssim 10$ au) accretes through the MRI-active surface layers, while a magnetically dead zone is formed at the disk midplane \citep{Gammie96,Simon+13}.
Although there may be quantitative uncertainty, this picture of layered accretion is considered generally valid, even in the presence of disk winds \citep{Armitage-Kley19}.
We modeled the magnetically dead zone using an effective and adaptive \cite{SS73} $\alpha$ parameter \citep{Bae+14,Kadam+19}:
\begin{equation}
\label{eq:alphaeff}
    \alpha_{\rm turb} = \frac{\Sigma_{{\rm MRI}} \, \alpha_{{\rm MRI}} + \Sigma_{{\rm dz}} \, \alpha_{{\rm dz}}}{\Sigma_{{\rm MRI}} + \Sigma_{{\rm dz}}},
\end{equation}
where $\Sigma_{{\rm MRI}}$ is the gas surface density of the MRI-active layer and $\Sigma_{{\rm dz}}$ is that of the magnetically dead layer so that $\Sigma_{\rm g} = \Sigma_{{\rm MRI}} +\Sigma_{{\rm dz}}$. Here, $\alpha_{\rm MRI}$ and $\alpha_{{\rm dz}}$ correspond to the strength of the turbulence in the MRI-active layer and the dead zone, respectively.
The $\Sigma_{\rm MRI}$ is obtained from the ionization fraction ($x$), which in turn is determined by solving the ionization balance equation
\begin{equation}
(1-x)\xi = \alpha_{\rm{r}} x^2n_{\rm{n}} + \alpha_{\rm{d}} xn_{\rm{n}},
\label{eq:ion}
\end{equation}
where $\xi$ is the ionization rate, $\alpha_{\rm{r}}$ is radiative recombination rate, $n_{\rm n}$ is the number density of neutrals, and $\alpha_{\rm{d}}$ is the total rate of recombination on to the dust grains \citep{DK14,DS87,Balduin+23}.
We consider the ionization by cosmic rays and radionuclides; however, effects of stellar far-ultraviolet and X-ray radiation are neglected since their contribution is typically significantly smaller {in the inner disk and limited to the upper disk atmosphere, which is not resolved in our thin-disk model \citep{IG99, Bergin07, PerezBecker-Chiang11}.}

In the simulations, $\alpha_{\rm MRI}$ is set to the canonical value of $0.01$, while lower values of $10^{-3}$ and $10^{-4}$ are also explored, the latter being consistent with recent observations of low viscosity in PPDs \citep{Flaherty+15,Dullemond+18,Muro-Aren+18,Flaherty+20,Franceschi+23}.
3D MHD simulations of FU Orionis type outbursts suggest that the $\alpha$ values in the inner disk region directly involved in the outburst can be much larger than that in the outer disk \citep[e.g.,][]{Zhu+20}.
During the MRI-type outbursts, the inner disk gets significantly hot and exceeds ionization temperature for alkali metals of $\approx 1000 $ K \citep{UN81,UN88}. 
The local thermal ionization then increases significantly and dominates the total ionization fraction.
Thus, we set $\alpha_{\rm MRI}$ to the maximum value of 0.1, when the local thermal ionization fraction exceeds $10^{-10}$. 
The larger value of the $\alpha$ parameter results in a better fit to the duration of FUor-type outbursts that is inferred from observations \citep{Zhu+09,VMHD20}.
The ionization threshold value is larger than that required for MRI activation in the midplane; however, it is justified since the upper regions of the disk are significantly more ionized from external irradiation \citep{Bai11,Desch-Turner15}.  
The method of transitioning $\alpha_{\rm MRI}$ as a function of ionization fraction is a better approximation of disk physics, as compared to the previous criterion in \cite{VMHD20}, where this switch was made for the entire innermost five au region.
Inside the dead zone, a small value of residual viscosity corresponding to $\alpha_{{\rm dz}}=10^{-5}$ is considered, which may arise from purely hydrodynamic turbulence driven by the Maxwell stress in the active layer \citep{Fleming-Stone03,OH11}.

The co-evolution of the magnetic field is implemented in the flux-freezing limit, while the non-ideal MHD effects \citep{Balbus-Terquem01,Kunz-Balbus04} are neglected due to prohibitively expensive computational costs.
However, while calculating the thickness of the MRI active surface layer above the midplane ($\Sigma_{{\rm crit}}$), we take into account Ohmic diffusivity, which dominate the inner disk region.
Equating the wavelength of the most unstable MRI mode to the gas scale height of the disk yields
\begin{equation}
    \Sigma_{{\rm crit}} = \left[\left(\frac{\pi}{2}\right)^{1/4}\frac{c^2m_{\rm e}\langle\sigma v\rangle_{\rm en}}{e^2}\right]^{-2}B_z^2 H_{\rm g}^3 x^2 \, ,
    \label{eq:DZ}
\end{equation}
where $e$ is the charge of an electron, $m_{\rm e}$ is the mass of an electron and $\langle\sigma v\rangle_{\rm{en}}$ is the slowing-down coefficient { \citep[see][for more details]{VMHD20}}. 
We note that $\langle\sigma v\rangle_{\rm{en}}$ is now updated to the empirically verified $10^{-9}\,\mbox{cm}^3\,\mbox{s}^{-1}$ \citep{Draine11,Das-Basu21}, instead of the previously used value of $10^{-7}\,\mbox{cm}^3\,\mbox{s}^{-1}$ that was based on a classical estimate \citep{Nakano84}.
The dead zone is present only when $\Sigma_{\rm g} > 2 \Sigma_{\rm crit}$, in which case $\Sigma_{\rm MRI}$ is set to $2\Sigma_{\rm crit}$.
If $\Sigma_{\rm g} \leq 2 \Sigma_{\rm crit}$, the disk is considered to be fully MRI active and there is no dead zone.
{ The MRI-active surface layer, as calculated with Eq. \ref{eq:DZ}, depends sensitively on the ionization balance, thermal balance and the magnetic field evolution in the disk. 
However, it results in a low-viscosity dead zone in the approximately innermost ten au of the disk (Eq. \ref{eq:alphaeff}), as anticipated by simulations \citep{Gammie96,Dzyurkevich10,VMHD20}.}
The evolution of $B_{\rm z}$ is computed by solving the advection equation and the planar components of magnetic field, $B_{\rm p}$, is evaluated as the solution of a magnetic analogue of the Poisson integral \citep{VMHD20}.
{ We note that the assumption of flux-freezing limit leads to significantly overestimated $B_{\rm z}$, which has consequences for the disk wind model.}

\subsection{Magnetic disk wind model}
\label{subsec:wind}

The magnetic disk wind model that we adopt is primarily based on the insights gained from the local MHD shearing box simulations of \cite{Bai13}, { which we modify for long-term disk evolution}.
The results of a series of shearing box simulations conducted at different radii can be stitched together to obtain a global picture of wind properties throughout the disk.
\cite{Bai13} derived the following fitting formulae for the radial dependence of wind mass loss rate (more accurately, rate of gas mass loss per unit area of the disk) and the wind-driven surface stress (i.e., the $z\phi$ component of Maxwell's stress tensor) for a protoplanetary disk (his Eqs. 9 and 10):  
\begin{equation}\label{eq:fitmw}
\frac{\dot{\Sigma}_{w,g}}{\rho_0c_s}\approx3.08\times10^{-5}
\frac{1}{N}
\bigg(\frac{r}{\rm au}\bigg)^{0.70}\bigg(\frac{\beta_0}{10^5 N}\bigg)^{-0.46} ,
\end{equation}
\begin{equation}\label{eq:fitTm}
\frac{T_{z\phi}^{\rm Max}}{\rho_0c_s^2}\approx1.16\times10^{-4}
\frac{1}{N}\bigg(\frac{r}{\rm au}\bigg)^{0.46}\bigg(\frac{\beta_0}{10^5 N}\bigg)^{-0.66}.
\end{equation}
Here, the quantities on the left hand side are normalized with respect to local midplane density ($\rho_0$) and sound speed, while $r$ is the stellocentric distance and $\beta_0$ is the ratio of gas to magnetic pressure at the disk midplane.
The normalization factor  
$ N = {{\Sigma_{\rm g}}/\Sigma_{\rm MMSN}} $
applies to a disk which diverges from minimum mass solar nebula (MMSN) density distribution.
With these two equations, one can implement a phenomenological prescription of disk winds; the $\dot{\Sigma}_w$ constitutes a sink term in the gas continuity equation, while a sink term in the momentum equation can be obtained from $T_{z\phi}^{\rm Max}$, accounting for the loss of mass and angular momentum through the winds, respectively.

Before Eqs. \eqref{eq:fitmw} and \eqref{eq:fitTm} could be used in a magnetohydrodynamic model, however, they need several modifications, and here we elaborate on these changes.
Firstly, the radial term in these formulae is essentially an MMSN radius, in the sense that it corresponds to the stellar mass of one $M_\odot$ ($r \equiv r_{\rm M_\odot}$). 
A shearing box is assigned a radial location in the disk based on its average angular velocity, $\Omega$.
Thus, to modify Eqs. \eqref{eq:fitmw} and \eqref{eq:fitTm} for an arbitrary stellar mass, the $(r_{\rm M_\odot}/ {\rm au})$ term needs to be modified so as to have the same angular velocity: 
\begin{equation}
\Omega^2 = \frac{GM_\star}{r^3} = \frac{GM_\odot}{r_{\rm M_\odot}^3},
\end{equation}
where $M_\star$ is the mass of the central star.
This implies that the radial dependence in Eqs. \eqref{eq:fitmw} and \eqref{eq:fitTm} should be replaced by
\begin{equation}
{r_{\rm M_\odot}}= r \bigg(\frac{M_\star}{M_\odot}\bigg)^{-1/3}
\end{equation}
in order to generalize these equations for a system with varying stellar mass.

Calculation of mass and angular momentum loss due to winds as prescribed by Eqs. \eqref{eq:fitmw} and \eqref{eq:fitTm} requires the disk plasma $\beta_0$, that is, the midplane ratio of gas to magnetic pressure. 
Since FEOSAD evolves the magnetic field, $\beta_0$ can be obtained throughout the disk (Sect. \ref{subsec:disk}). 
{ We note that $\beta_0$ is defined differently in \cite{Bai13} (Heaviside-Lorentz units, $\beta_{\rm wind} = {2 P_0}/{B_z^2}$) and in the hydrodynamic model of \cite{VMHD20} (Gaussian units, $\beta_{\rm MHD} = {8 \pi P_0}/{B_z^2}$), where $P_0$ and $B_z$ are midplane pressure and vertical magnetic field, respectively.} 
In the context of calculating the wind quantities, $\beta_0 = \beta_{\rm wind}$ is used.
When the mass loss rate due to the winds over the entire disk is calculated by post-processing FEOSAD simulations (i.e., Eq. 13 applied to surface density of a gravitoviscous simulation), we found that this integrated mass loss rate was consistently about two orders of magnitude larger than the accretion rate onto the star.
Such a high rate of mass loss is unrealistic and not sustainable, as the entire disk will be lost within $10^4$ yr.

There are two reasons for a substantially large wind mass loss rate while using \cite{Bai13}'s equations as they stand.
Firstly, we found that the local $\beta_0$ in FEOSAD simulations is generally much lower than the values widely accepted in the literature.
Measuring the magnetic field strength in protoplanetary disks remains challenging, although it is expected to be strongly subthermal, with a typical T Tauri disk exhibiting $\beta_0 \lesssim 10^4$ \citep{Vlemmings+19,Lesur21b}.
Similarly large values of $\beta_0 \sim 10^2 - 10^8$ are often considered in shearing box simulations \citep[e.g.,][]{Bai13, Simon+18}, { which are in alignment with global disk simulations that include non-ideal MHD effects \citep{Masson+16,Xu-Kunz21,Mauxion+24}.} 
However, in FEOSAD simulations, $\beta_0$ is consistently lower, indicating the presence of a strong magnetic field.
This discrepancy occurs due to the ideal MHD limit for evolving the magnetic field, where the accreting gas tends to drag the field with it, {increasing $B_{\rm z}$ and thus} decreasing local values of $\beta_0$.
Secondly, the shearing box-simulations may severely overestimate the mass loss rate, as the mass loss rate consistently decreases with an increase in the box height in the direction normal to the disk midplane. 
This term is highly uncertain and \cite{Bai13} proposed an additional correction proportional to $H_g/r$, in order to account for this dependence on the finite size of the shearing box. 
In our model, we attempt to remedy the large magnetic fields in the MHD simulations via a constant correction factor, 
\begin{equation}
{ C_\beta} = 1000,
\end{equation}
that is multiplied with $\beta_0$ while calculating the winds.
This factor reflects the limitations of current simulations and may be dropped in the future when the non-ideal MHD equations are solved. 
{ Parameter $C_\beta$ is constrained such that $C_\beta \times \beta_0 \sim 10^{4}$ over majority of the disk, which is close to the expected values of $\beta_0$ in protoplanetary disks. However, we note that there is relatively large amount of uncertainty in this estimate from both observational and theoretical side.}
In order to account for the finite size of the shearing box, we include a term proportional to the local disk aspect ratio in the equation for wind mass loss $\dot{M}_w$: 
\begin{equation}
{C_{\rm Box}} = {C_H} \frac{H_{\rm g}}{r},
\label{eq:cbox}
\end{equation}
where ${C_H}$ is a correction factor with the default value of 0.5.
When both the ${\rm C_\beta}$ and ${\rm C_{Box}}$ terms are considered, the integrated disk wind mass loss rate is about the same order of magnitude as the accretion rate onto the star.  
The correction ${\rm C_{Box}}$ only applies to the mass loss rate and not the surface stress, as the latter is well constrained in the shearing box simulations.
{ This parameter is constrained by the total mass loss rate from the winds such that the latter is of the order of the accretion rate onto the central protostar \citep{Pascucci+23}.}
Concerning the total mass loss from the system, we additionally assume formation of a high velocity jet close to the star in all models, including those with only gravitoviscous evolution. 
The jet is formed concurrently with the star and it lies inside the central sink cell, and 10\% of the accreting mass is assumed to be ejected via the jet \citep{Hartigan+94,Bally16}.

The phenomenon of episodic accretion plays an important role in the long-term evolution of a protoplanetary disk. 
In particular, FEOSAD simulations show luminous MRI outbursts, which are analogous to FUor type eruptions observed in young stellar objects \citep{Hartmann-Kenyon96, Kadam+20, VMHD20}. 
These eruptions are thought to be accompanied by enhanced ejections, as indicated by observations of variability in outflow signatures, manifesting as clumpy structures in jets and outflows \citep{Reipurth89,VE18,Fischer+23}. 
Approximately 10\% of the accreting mass may be expected to be lost via molecular outflows that are associated with the low velocity, wide angle winds \citep{Audard+14,Fernando+23}.
For our model, we interpret the enhanced outflows during outbursts as magnetic disk winds that are linked to the increased far-ultraviolet (FUV) luminosity of the central accreting star \citep{Bethell-Bergin11}.
The FUV ionization near the disk surface is essential for making the gas in the disk well-coupled to the magnetic field \citep{Gorti-Hollenbach09}. 
The efficiency of wind transport increases with the increase in FUV penetration depth, $\Sigma_{\rm FUV}$, which is proportional to incident radiation \citep{Simon+13,Bai16}. 

In the shearing box simulations that result in the relations specified in Eqs. \eqref{eq:fitmw} and \eqref{eq:fitTm}, { the FUV penetration depth is considered to be a constant ($\Sigma_{\rm FUV}=0.03$ ${\rm g \,  cm^{-2}}$, in accordance with \cite{PerezBecker-Chiang11}) and effects of changing $\Sigma_{\rm FUV}$ are not parameterized.}
{ To the first approximation, when $\Sigma_{\rm FUV}$ is increased by an order of magnitude, the wind mass loss increases by about the same factor, while the wind stress also increases, but only by a factor of a few \citep{BS13}.
This guides the calibration of the exponents in the following correction terms that are introduced in order to account for varying FUV luminosity}in the wind mass loss rate and surface stress equations, respectively:
\begin{equation}
{{\rm C}_{{\rm FUV}, \dot{M}_w}}= \frac{1}{2}\bigg(\frac{L_T}{L_\odot}\bigg)^{0.8},
\label{eq:cfuvm}
\end{equation}
\begin{equation}
{{\rm C}_{{\rm FUV}, T_{z \phi}}}= \frac{1}{2}\bigg(\frac{L_T}{L_\odot}\bigg)^{0.5}.
\label{eq:cfuvt}
\end{equation}
Here, $L_{T}$ is total, that is, the accretion plus stellar luminosity of the system.
The exponent for the ${\rm C}_{{\rm FUV}, \dot{M}_w}$ term is chosen to be marginally lower than unity, since a larger value results in a significant decrease in the wind velocity during an outburst (see Appendix \ref{app:windvel}).
We note that this dependence incorporates several assumptions implicitly, for example, the FUV penetration depth is proportional to the central FUV luminosity, which in turn is proportional to the total luminosity of the accreting star, along with the exact exponents of the wind mass loss and stress.
Additionally, the dusty winds can self-shield the FUV irradiation; the winds from the inner regions can absorb FUV photons and interfere with the wind launching from the remainder of the disk \citep{Panoglou+12,Rab+22}.  
With the exponents in the above equations being less than unity, self-shielding is inherently assumed in the model.
We neglect any additional uncertainties, for example, effects of photochemistry in the upper layers on the ionization \citep{Bergin+03,Woitke2009}.
Inclusion of these terms (Eqs. \eqref{eq:cfuvm} and \eqref{eq:cfuvt}) results in an increased wind mass loss rate during outbursts that is approximately an order of magnitude less than the accretion rate onto the star.
A caveat of this approach is that the radiation takes a finite time to reach the disk, which should cause a delay in producing magnetic winds.
However, this delay is a small fraction of the dynamical time at a given radius, and thus, it can be safely neglected.

As the winds originate centrifugally from a protoplanetary disk, it is essential to determine the precise extent of this disk.  
During the collapse of molecular cloud to form a star, a gas parcel of specific angular momentum $j$ will be accelerated toward the central star in a parabolic orbit.
This orbit intersects the plane of the disk at
\begin{equation}
    r_{\rm eq} = \frac{j^2}{G M_{\rm enc}},
\end{equation}
where $M_{\rm enc}$ is the mass enclosed within the orbit of the gas parcel \citep{Dominik15}.
A gas parcel can be part of the centrifugal disk only if it has sufficient specific angular momentum to resist the inward fall.
This criterion implies that material at a given radius $r$ forms a centrifugal disk, if the ratio $r/r_{\rm eq}$ is less than unity.
In a protoplanetary disk, however, the gas experiences additional pressure support, and hence, it moves with a sub-Keplerian velocity.
Thus, for calculating the outer boundary of the centrifugal disk, we use a less stringent criterion of $r/r_{\rm eq} < 1.2$.
The disk winds are active only within this region and are completely inactive outside.
In addition, the disk winds are turned off below a gas surface density threshold of $0.1 {\rm g \,cm^{-2}}$, in order to limit unintended consequences of the winds at large distances.
We note that due to the diminishing gas surface density, the wind mass loss rate and the surface stress decrease with radius and become progressively insignificant.
We obtain a single ``centrifugal radius" ($R_{\rm cf}$) by azimuthally averaging the extent of the centrifugal disk at a given time.

The action of the disk winds needs to be taken into account self-consistently in the hydrodynamic equations that are solved in FEOSAD.
For this, the equations that need to be modified are - mass continuity equation for gas and small dust (Eqs. \eqref{eq:cont} and \eqref{eq:contDsmall}), momentum equation (Eq. \eqref{eq:mom}) and energy transport equation (Eq. \eqref{eq:energy}).
For solving the MHD equations, FEOSAD uses operator splitting technique, which divides problem into two partial substeps- the source and the transport step \citep[e.g.,][]{Stone-Norman92}.
During the numerical calculations, the corrections due to wind mass loss and surface stress ($\dot{\Sigma}_{w}$ and $T_{z \phi}^{\rm Max}$) are computed in each cell at the beginning of the source step.
Corrections to gas and small dust surface density as well as energy are done at the end of the source step, but before the transport step begins.
A fraction, $d\Sigma_{w,g}$, is removed from the gas surface density such that 
\begin{equation}
    d\Sigma_{\rm w,g} = \dot{\Sigma}_{\rm w, g}  dt,
\end{equation}
where $dt$ is the current time step. 
Even though the winds originate well above the midplane, aerodynamically coupled dust grains can get entrained in the outflows and leave the disk \citep{Miyake+16,Booth-Clarke21}.
Hence, an amount proportional to the small dust-to-gas mass ratio is subtracted from the small dust surface density, which is 
\begin{equation}
    d\Sigma_{\rm w,d} = \frac{\Sigma_{\rm d,sm}}{\Sigma_{\rm g}}  {d \Sigma_{\rm w, g}} \, .
\end{equation}
The surface stress $T_{z \phi}^{\rm Max}$ has the units of momentum per unit area per unit time and it is directly subtracted from the $\phi$ component of the momentum equation.
The internal energy equation is balanced in such a way that the local gas temperature (energy per unit mass) is unchanged when the effect of the wind is applied, which results in the correction term
\begin{equation}
d e_{\rm w} = \frac{d \Sigma_{\rm w,g}} {\Sigma_{\rm g}} e.
\end{equation}
In order to prevent a sudden, large amount of mass loss during a time step, the mass lost due to the wind is capped at 5\% of the gas mass contained within the cell.
In such a case, the disk  $T_{z\phi}^{\rm Max}$ is capped such that the velocity of the escaping wind remains unchanged (see Appendix \ref{app:windvel} for details).

To summarize, our model assumes that the magnetic winds are symmetrical across the disk midplane on both sides. 
We also assume that the wind properties can be expressed as power law functions of the local quantities and stellar parameters.
The final modified fitting formulae describing wind mass loss rate and surface stress are
\begin{equation}\label{eq:Mw1}
\frac{\dot{\Sigma}_{w,g}}{\rho_0c_s}=
C_{\dot{M}}
\frac{{ C_H} H_{\rm g}/r}{N} 
\bigg(\frac{M_\star}{M_\odot}\bigg)^{\!-0.23}
\bigg(\frac{r}{\rm au}\bigg)^{\!0.7}
\bigg(\frac{{C_\beta}\beta_0}{10^{\,5}N}\bigg)^{\!-0.46} {\bigg(\frac{L_T}{L_\odot}\bigg)^{\!0.8}},
\end{equation}
\begin{equation}\label{eq:Tm1}
\frac{T_{z\phi}^{\rm Max}}{\rho_0c_s^2}=
C_{T_{z\phi}}
\frac{1 }{N}
\bigg(\frac{M_\star}{M_\odot}\bigg)^{-0.15}
\bigg(\frac{r}{\rm au}\bigg)^{0.46}\bigg(\frac{{C_\beta}\beta_0}{ 10^5 N}\bigg)^{-0.66}\ 
{\bigg(\frac{L_T}{L_\odot}\bigg)^{0.5}},
\end{equation}
where $C_{\dot{M}}=1.54\times10^{-5}$ and $C_{T_{z\phi}}=5.8\times 10^{-5}$ are obtained by consolidating all other constants in the respective equations. 
With this framework of magnetic disk winds, all of the exponents in Eqs. \eqref{eq:Mw1} and \eqref{eq:Tm1} are informed by the shearing box simulations.
The only ``free" parameters are $C_H$ and $C_\beta$, which arise from the discrepancy with respect to non-ideal MHD effects and uncertainties in the wind mass loss rate, respectively.  
We constrain these parameters with the help of observations in Sect. \ref{subsec:obscomp}.  
We note that with the assumption of spatially uniform mass to flux ratio, $\lambda=\Sigma_{\rm g}/B_{\rm z}$, $\dot{\Sigma}_{w,g} \sim \Sigma_{\rm g} c_{\rm s}$ and $T_{z\phi}^{\rm Max} \sim \Sigma_{\rm g}^{5/4} c_{\rm s}^{2/3}$, which explains how the disk wind properties vary with the local quantities.

\subsection{Interface to {\sc ProDiMo}}
\label{subsec:prodimo}

The disk structures computed by FEOSAD are passed to the radiation thermo-chemical disk model {\sc ProDiMo} \citep{Woitke2009,Woitke2016,Woitke2023} for post-processing.
This way, we can predict the optical appearance of these disks at millimeter wavelengths in the continuum and in CO lines to simulate ALMA Band 6 and Band 7 observations, and read off the apparent sizes of the disks as the radii that encircle 90\% of the respective continuum and line fluxes.
{ We choose the inclination to be face-on to avoid the necessity to de-project while producing ray-traced continuum images and CO line maps.}

The main challenge for this interface is that FEOSAD uses 2D polar coordinates $(r,\phi)$, considering vertically integrated quantities, whereas {\sc ProDiMo} uses 2D cylinder coordinates $(r,z)$, assuming the disk to be axisymmetric. We have therefore developed the following scheme. First, the FEOSAD disk structure at a given time is azimuthally averaged. 
A PPD structure is not strictly axisymmetric; however, the azimuthal asymmetries increasingly become smoothed out as the disk evolves and the GI diminishes. 
Second, the following quantities are passed to {\sc ProDiMo}: the gas column density $\Sigma_{\rm g}$, the column density of small dust $\Sigma_{\rm d,sm}$, the column density of grown dust $\Sigma_{\rm d,gr}$, the viscosity parameter $\alpha_{\rm turb}$, the scale height $H_{\rm g}$, and the maximum grain radius $a_{\rm max}$. 
All these quantities are passed to {\sc ProDiMo} as vectors on approximately 200 log-equidistant radial grid points used by FEOSAD.  
In addition, the stellar luminosity $L_\star$, the effective temperature of the star $T_{\rm eff}$ and the stellar mass $M_\star$ are passed as scalars.
Third, {\sc ProDiMo} sets up the 2D structure of the gas disk in $(r,z)$ on the provided radial grid points, while adding a few additional grid points by extrapolation near the inner rim so that the penetration of stellar light into the disk is traced correctly. 
We then use the radius-dependent scale height $H_g(r)$ to set up the vertical gas density structure $\rho(r,z)$ at each radius using a Gaussian function.

The dust grains in FEOSAD have size distribution between a fixed minimum size $a_{\rm min}\!=\!0.005\,\mu$m and a variable maximum size $a_{\rm max}$, while assuming a continuous power-law size distribution of index –3.5.  
In each vertical column, {\sc ProDiMo} sets up the unsettled grain size distribution function using 100 size bins, and then the grains are settled according to the prescription of \citet{Riols2018} in each bin with the settling parameter $\alpha_{\rm turb}$ \citep{Woitke2023}.
We use the DIANA standard dust opacities \citet{Woitke2016}, assuming the grains to be composed of 60\% silicate, 15\% amorphous carbon, with 25\% porosity { and without any ice. 
It is possible that ice formation can locally increase the dust opacities by more than a factor of 100, especially at UV to mid-IR wavelengths \citep{Arabhavi2022}. 
However, the ice exists only deep in the disk ($A_V\!\gtrsim\!10$), where it is protected from UV photons, that is, where the refractory dust is optically thick. Therefore, ice opacities do not have a strong influence on the resulting dust temperature structure and the spectral appearance of the disk concerning CO sizes and millimeter fluxes. Although some uncertainties remain when vertical mixing is included in the models \citep{Woitke2022}, in which case the grains can be mixed up faster than photodesorption can destroy the ice.}
Based on the settled dust structure, stellar and interstellar irradiation, and internal viscous heating, {\sc ProDiMo} performs continuum radiative transfer calculations.
This results in the disk internal dust temperature structure $T_{\rm d}(r,z)$ and produces ray-traced continuum images and CO line maps using a polar grid in the image plane with $225\times72$ segments.

We use the large DIANA standard chemical network \citep{Kamp2017} for the thermo-chemical modeling part, which has 235 chemical species, with reaction rates mostly taken from the UMIST\,2012 database \citet{McElroy2013}, with added H$_2$ formation on grains, a simple freeze-out and desorption ice chemistry, X-ray processes including doubly ionized species \citet{Aresu2011}, excited molecular hydrogen, and polycyclic aromatic hydrocarbons in five different charging states; altogether there are 4832 reactions. 
The most relevant chemical processes for this paper are the CO photo-dissociation in the upper disk regions, and the CO freeze-out in the outer midplane regions. For simplicity we solve for the chemical concentrations of all ice and gas phase species in kinetic chemical equilibrium.
Recent code improvements concerning escape probability theory and photo-rates are explained in \cite{Woitke2023}.

The viscosity parameter $\alpha_{\rm turb}(r)$ is also used for computing the viscous heating in {\sc ProDiMo}.  
Here we use a 2D diffusion solver in the optically thick core of the disk as described in the appendix of \citet{Oberg2022} and use the following formula for the local dust heating \citep{APIA02},
\begin{eqnarray}
  \Gamma_{\rm col}(r) &=&
  \frac{9}{8}\,\alpha_{\rm turb}\,H_{\rm g}^2\,\Sigma_{\rm g}\,\Omega^3 ,\\
  \Gamma(r,z) &=& \Gamma_{\rm col}(r)
  \frac{\rho_{\rm d}(r,z)}{\int_0^\infty \rho_{\rm d}(r,z')\,dz'} \ ,
\end{eqnarray}
where $\Gamma_{\rm col}(r)\,\rm[erg\,s^{-1}\,cm^{-2}]$ is the viscous heating rate of a column and $\Gamma(r,z)\,\rm[erg\,s^{-1}\,cm^{-3}]$ is the local viscous heating rate per volume $\big(\int_0^\infty\Gamma(r,z')\,dz'=\Gamma_{\rm col}(r)\big)$. 
All {\sc ProDiMo} models have been performed with git version {\tt 4fee3902} from 2023/06/19.

\subsection{Initial conditions and model parameters}
\label{subsec:init}

The MHD simulations in FEOSAD start with the gravitational collapse phase of a starless molecular cloud core.
The initial mass distribution as well as angular velocity is consistent with axisymmetric core collapse, where the angular momentum remains constant and magnetic fields are expelled due to ambipolar diffusion.
The collapse of the core continues such that a star is formed within the central sink of the computational domain.
A surrounding protostellar disk simultaneously forms while the envelope is infalling and the system continues to evolve through the embedded phase.
{Detailed description of the initial structure and angular velocity profile can be found in \cite{Basu97} and \cite{VMHD20}.}   
Since the focus of this study is on the effects of the magnetic winds on the disk formation and its evolution, all simulations start with identical initial conditions, which are summarized in Table \ref{table:params}. 
The cloud core of 0.83 $M_\odot$ forms the initial gas mass reservoir and results in a young star of mass between $0.5 - 0.7 M_\odot$, depending on the efficiency of the particular model. 
The ratio of kinetic to gravitational energy is set to $E_{\rm rot}/ E_{\rm grav} = 0.23 \%$, consistent with the observations of prestellar cores \citep{Caselli+02,Li+12}.
The disks formed with these conditions are gravitationally unstable, while being stable to fragmentation and clump formation \citep{Vorobyov13}.
These particular initial conditions are chosen because formation of disks in similar conditions is extensively studied with FEOSAD simulations \citep[e.g.,][]{Kadam+20,VMHD20}.
Additionally, we compare our results with the ALMA survey of low mass star forming regions, wherein the stellar mass of the sample centers at M0 type star of mass $\approx 0.5$ $M_\odot$ \citep{Ansdell+16,Manara+23}.

The initial bulk dust-to-gas mass ratio, $\zeta_{\rm d2g}$, is set to the canonical interstellar medium (ISM) value of 1\% and all the dust mass is in the fully coupled submicron-sized small particles. 
The initial cloud core gas temperature is 20 K, which was also the temperature of the background radiation \citep{KE12}. 
A uniform background magnetic field of $10^{-5}$ G normal to the plane of the disk is also assumed \citep{Liu+21}. 
The mass-to-flux ratio ($\lambda_c = \Sigma_g/ B_z$, in the units of the critical value $1/2\pi G^{1/2}$) in dense star-forming cores is observed to be $\lambda_c \simeq 2$ \citep{Crutcher12}, with a cloud or core being supercritical for values of $\lambda_c$ above unity \citep{MS76,Nakano-Nakamura78}.
A commensurate value of $\lambda_c =2$ was used for our earlier investigations; however, some flux should be lost during the collapse phase due to ambipolar diffusion \citep{Shu+87,Dapp+12}. 
Since FEOSAD does not include non-ideal MHD effects, we can justify a somewhat larger value of $\lambda_c = 5$ and thus partially compensate for the model limitations.
The resolution of all simulations is $184 \times 128$ in polar coordinates ($r, \phi$), with a logarithmic spacing in the radial direction and uniform spacing in the azimuthal direction at each radius. 
The increased resolution in the $r$-direction is chosen to minimize the elongation of grid cells, preventing potential numerical artifacts. 
The convergence of FEOSAD code at different spatial resolutions has been previously confirmed \citep[e.g.,][]{Kadam+21} and the current resolution sets $dr =2.6 \times 10^{-2}$ au at the inner computational boundary at 0.53 au.
The location of this inner boundary or sink cell excludes the actual inner disk radius of a PPD occurring at $\approx 0.1$ au, which is determined by the magnetospheric accretion or dust sublimation radius \citep{Dullemond-Monnier10,Hartmann+16}.
However, the marginally larger boundary relaxes the Courant condition for computational feasibility of the simulations \citep{CFL28}, while being sufficient for capturing important dynamical phenomena occurring in the inner disk, such as episodic accretion.
A carefully constructed inflow-outflow boundary condition is implemented at the inner sink cell in order to prevent the artificial drop in the gas surface density \citep{VDust18,Kadam+22}.

\begin{table}
\caption{Initial parameters common across simulations.}
\label{table:params}
\begin{tabular}{p{3cm}p{3cm}}
\hline
Model parameter &   Value\\
\hline
$M_{\rm core}$  &   0.83 $M_\odot$\\
$E_{\rm rot}/E_{\rm grav}$         &   $0.0023$\\
$\lambda_c$       &   5\\
$T_{\rm init}$   &  20 K \\
$\zeta_{\rm d2g}$  &    0.01   \\
$r_{\rm in}$     &  0.53 au \\
$r_{\rm out}$    &  4027 au\\
\hline
\end{tabular}
\end{table}

\begin{table*}
\caption{List of simulations.}
\label{table:sims}
\begin{tabular}{l}
\hline
\begin{tabular}{p{1.7cm}p{1.4cm}p{1.4cm}p{1.4cm}p{7.5cm}}
 Model  &  $\alpha_{\rm MRI}$ & $C_H$ & $C_\beta$  & \hspace{1.5cm} {Significance}  \\
\end{tabular}\\ \hline
\begin{tabular}{l}
$\kern-\nulldelimiterspace\left.
\begin{tabular}{p{1.4cm}p{1.4cm}p{1.4cm}p{1.4cm}p{5.8cm}}
\simname{GV-2}   &   $10^{-2}$  &  0.5 &   1000 & Fiducial high $\alpha$ \\
\simname{GV-4}   &   $10^{-4}$ & 0.5 &   1000 & Fiducial low $\alpha$ \\
\end{tabular}\right\}$ No winds$^1$ 
\\ 
\end{tabular}  
\\ \hline
\begin{tabular}{l}
$\kern-\nulldelimiterspace\left.
\begin{tabular}{p{1.4cm}p{1.4cm}p{1.4cm}p{1.4cm}p{5.8cm}}
\simname{WI-2}   &    $10^{-2}$  & 0.5 &   1000 &  High $\alpha$  \\ 
\simname{WI-4}   &   $10^{-4}$ & 0.5 &   1000 &  Low $\alpha$  \\ 
\simname{WI-3}   &    $10^{-3}$  & 0.5 &   1000 &  Intermediate $\alpha$ \\ 
\simname{WI-3a}   &   $10^{-3}$ & 1.0 &   5000 &  Intermediate $\alpha$, adjusted wind parameters\\ 
  \end{tabular}\right\}$ With disk winds
\end{tabular}  
\\
\hline
\end{tabular}\\
\raggedright $^1$ Wind quantities are only calculated and not included in the evolution equations. \\
\end{table*}

In this study, we present results from six simulations as listed in Table \ref{table:sims}.
The fiducial simulations, \simname{GV-2} and \simname{GV-4}, evolve via the combined action of gravitational and viscous torques. 
The wind quantities in these models are calculated; however, no corrections are made to the disk evolution equations.
The value of background $\alpha_{\rm MRI}$ in \simname{GV-2} is 0.01. 
As discussed in Sect.\,\ref{sec:intro}, protoplanetary disks may exhibit significantly lower value of turbulent $\alpha$ and this possibility is explored in model \simname{GV-4}, with $\alpha_{\rm MRI}=10^{-4}$.
Model \simname{WI-2} is identical to \simname{GV-2}, wherein the disk winds are taken into account self-consistently.
The gravitoviscous forces are active in all disk models, including those where the magnetic winds are included.  
In the case of \simname{WI-4}, winds are considered with a lower value of $\alpha_{\rm MRI}$, and thus, it is otherwise identical with \simname{GV-4}.
Model \simname{WI-3} considers winds with the intermediate value of $\alpha_{\rm MRI}=10^{-3}$. 
In the case of \simname{WI-3a}, the wind parameters $C_H$ and $C_\beta$ are adjusted, in order to demonstrate how a better fit to the observations can be obtained.
With this limited parameter space study some of the most important possibilities in the disk wind scenario are explored.

\section{Results}
\label{sec:results}

\subsection{Global picture of disk evolution}
\label{subsec:global}

In this section we describe the process of disk formation and its evolution on a global scale.
FEOSAD simulations with similar initial conditions, model parameters and disk physics have been studied previously in detail in our group. 
In particular, the fiducial simulation \simname{GV-2} is analogous to model 2 in \cite{VMHD20} and \cite{Kadam+22}. 
The motivation behind reanalyzing the gravitoviscous evolution of \simname{GV-2} and \simname{GV-4} is to confirm that our current disk model behaves as expected with the inclusion of incremental improvements and establish it as a benchmark in order to study and contrast the effects of including the action of magnetic winds.

\begin{figure*}
\centering
  \includegraphics[width=\textwidth]{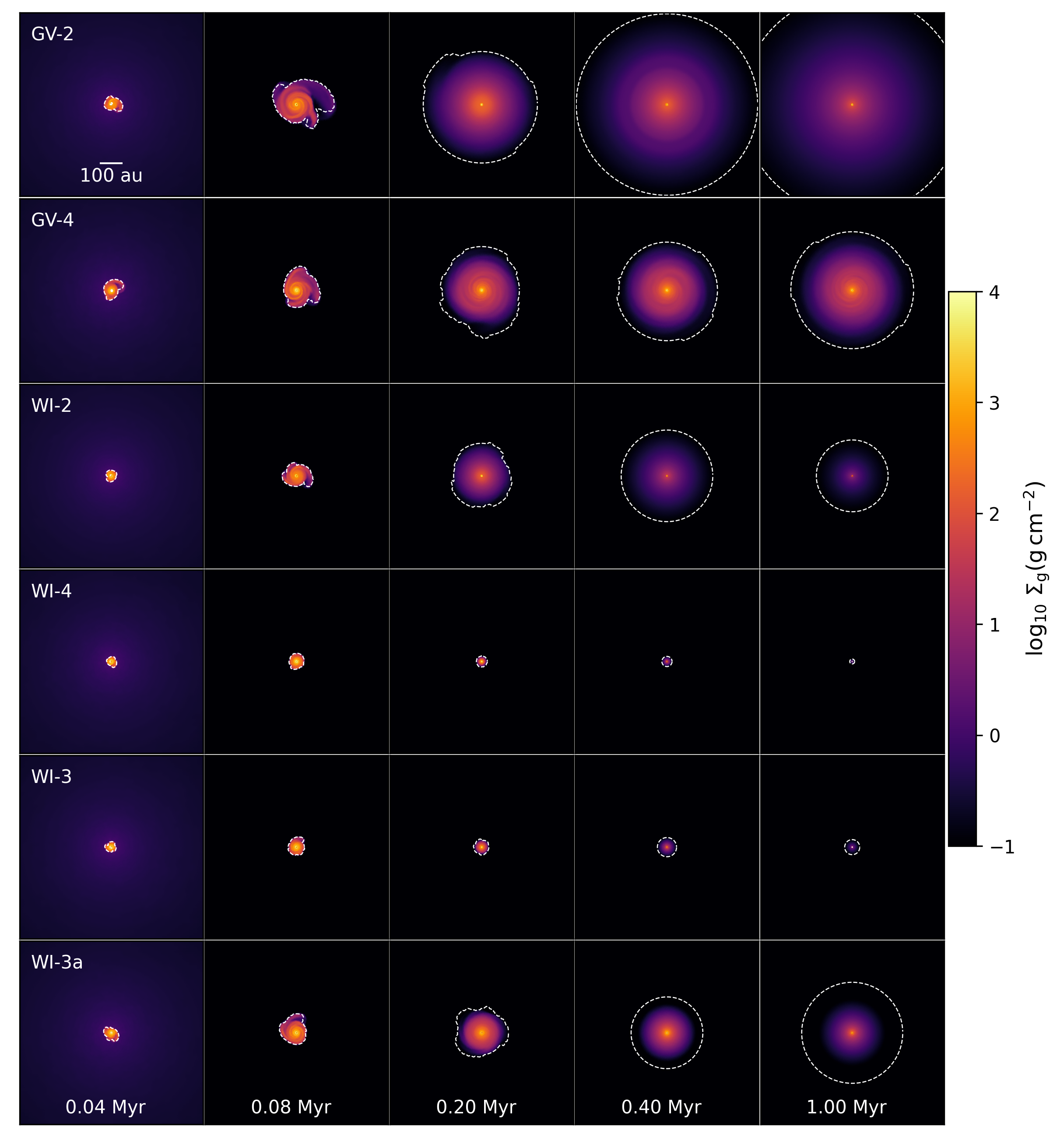}
\caption{Evolution of the gas surface density for the protoplanetary disk models showing the global picture. The white contours show the extent of the centrifugal disk. We note that the time intervals are not uniform and are chosen to depict viscous spread of the disks. (See Table \ref{table:final} for a summary of the properties of the resulting systems.)}
\label{fig:global}
\end{figure*}

As soon as a simulation begins, the initial prestellar cloud core starts to collapse, since it is constructed to have a supercritical mass-to-flux ratio and is gravitationally unstable. 
The time is measured from the onset of collapse, that is, the start of the simulation. 
In the initial stages of the collapse, surface density increases obeying a nearly self-similar law \citep{BM94, Basu97}, while the rotation profile remains rigid within about 50 au. 
As the cloud continues to contract, the mass of the gas in the sink cell exceeds 0.02 ${\rm M_\odot}$, leading to the assumption of second collapse due to dissociation of hydrogen molecule \citep{Larson69}.
A protostar thus forms at the coordinate center at this time, with a point-like gravitational potential.
The stellar luminosity now illuminates the disk and the gas temperature in the innermost regions exceeds 600 K due to both stellar irradiation and compressional heating. 
The infalling gas in the innermost regions soon starts to decelerate while approaching the star and a centrifugal disk larger than the sink cell forms at about 0.022 Myr.
This initial evolution before the formation of the circumstellar disk is almost identical across all simulations.

Figure~\ref{fig:global} depicts the gas surface density of the disk models as listed in Table \ref{table:sims}, in the inner $800 \times 800$ au box, at different instances of time.
We note that the intervals between these instances are not uniform, but increase as the disk evolves, so as to highlight the viscous spreading of the disk.
The dashed line depicts extent of the centrifugal disk at a given time, as specified in Sect.\,\ref{subsec:wind}.
Consider the first two fiducial gravitoviscous models, \simname{GV-2} and \simname{GV-4}.
The general evolution of the two models looks similar at this scale.
The early stages are dominated by GI, as the disk mass is a large fraction of that of the central protostar.
This phase is characterized by formation of large-scale spirals, which are seen as azimuthal asymmetries in the gas surface density. 
With the current choice of initial conditions, formation of self-gravitating clumps is not expected. 
As the disk continues to evolve, its viscous spread is clearly seen in the increasing centrifugal radius, which marks the theoretical outer extent of the disk. 
Similar to \cite{Kadam+20}, we define the Class 0/I boundary as the time when the mass of the star-disk system exceeds half the mass of the initial cloud core.
The Class I/II boundary is considered to be the time when the envelope accretion rate drops below $10^{-8}\, {\rm M_\odot yr^{-1}}$.
In Fig. \ref{fig:global}, the first two frames capture the disks in the embedded Class 0 and Class I stages, respectively.
In the last three frames, a disk can be considered as a T Tauri object of Class II.
These class boundaries occur at roughly the same time in all simulations. 

The major difference between models \simname{GV-2} and \simname{GV-4} at this global scale is the extent of the disk, especially during the Class II phase. 
\simname{GV-2} uses the value of $\alpha_{\rm MRI}$ as $0.01$, while that for \simname{GV-4} is $10^{-4}$.
We note that $\alpha_{\rm turb}$ can achieve a much lower value in the dead zone (see Eq. \ref{eq:alphaeff}).
However, outside of the dead zone ($\gtrsim 10$ au), the disk maintains the maximum possible value of $\alpha_{\rm turb} = \alpha_{\rm MRI}$.
The higher viscosity in the bulk of the disk causes \simname{GV-2} to evolve on a faster viscous timescale and show a larger viscous spread.
The azimuthal asymmetries and disk substructure in the outer regions also get smoothed out with higher viscosity, producing a nearly perfect axisymmetric disk.
The centrifugal radius of the disk is several hundred au, which is comparable with the gas radii of some of the largest T Tauri disks such as DM Tau or IM Lup \citep{Dartois+03,Cleeves+16}.
On the other hand, the disk in \simname{GV-4} shows a lesser viscous spread.
Lower viscosity in the disk is unable to efficiently get rid of angular momentum and transport the gas inward. 
Thus, the gas remains in the disk for a longer time and forms a smaller ($\approx 250$ au) but relatively massive disk, which shows spiral substructure even at 0.8 Myr.
The disk radii and masses are discussed quantitatively in Sect. \ref{subsec:obscomp}, with respect to synthetic observations.

The next four rows show models, \simname{WI-2}, \simname{WI-4} and \simname{WI-3}, and \simname{WI-3a}, which include the disk winds self-consistently in the evolution equations. 
As specified in Table \ref{table:sims}, the first three models have an identical prescription of magnetic disk winds, while \simname{WI-3a} considers adjusted wind parameters. 
Consider the evolution of model \simname{WI-2}; the disk shows asymmetry and spirals during the initial stage of evolution.
During the Class II stage, the disk is much more compact as compared to analogous gravitoviscous model \simname{GV-2} and also smaller than the low viscosity model \simname{GV-4}. 
In the last frame, the disk radius shrinks marginally, instead of expanding, resulting in a disk radius of $\approx 150$ au.
A purely viscous disk transports the angular momentum radially outward, redistributing it within the disk, which leads to a gradual expansion. 
Similarly, the gravitational torques work within the disk, but unlike viscosity, gravitational forces are inherently non-local and work over larger distances via non-axisymmetric perturbations and spiral density waves \citep{Kratter-Lodato16}.
With winds, the disk evolution is driven by removal of angular momentum instead of its transport leading to so called ``advective disk'', where the disk size tends to contract with time \citep{Alessi-Pudritz22}. 
In our models, the exact evolution of the disk size depends on the balance between gravitoviscous forces, which work toward expanding the disk and magnetic winds, which oppose this spread.

The evolution of the low and intermediate viscosity models, \simname{WI-4} and \simname{WI-3}, is peculiar; we obtain centrifugal disks that are very compact.
In the case of \simname{WI-4}, the disk extends to a maximum size of $\approx 25$ au during the Class I phase. However, it soon shrinks to a much smaller size of $\approx 6$ au, which is unusually small for a Class II disk.
The disk of model \simname{WI-3} achieves a maximum size of 45 au and then gradually shrinks to about $25$ au.
Although disk sizes of tens of au are well within the observational expectations for a gas disk, we later show in Sect. \ref{subsec:obscomp} that both models \simname{WI-4} and \simname{WI-3} produce dust disks that are unreasonably small in comparison with observations.
The trend with respect to disk size across different $\alpha_{\rm MRI}$ values confirms our hypothesis of the opposing action of disk winds and gravitoviscous expansion.
With the current prescription of winds, their action is significantly stronger as compared to the viscous spread produced at low $\alpha_{\rm MRI}$ values of $10^{-4} -10^{-3}$.

With \simname{WI-3a}, we explore the possibility of changing the wind strength by adjusting the wind-related parameters $C_H$ and $C_\beta$.
For this model, a larger value of $C_H$ results in a more efficient mass loss. The larger value of $C_\beta$ used implies weaker winds, as it results in a decrease in both the wind mass loss and the surface stress (see Table \ref{table:sims} and Eqs. \eqref{eq:Mw1}, \eqref{eq:Tm1}).
As seen in the last row of Fig. \ref{fig:global}, the disk in \simname{WI-3a} achieves a larger size as compared to the other two wind models at low viscosity, 
while also reaching a maximum size that is smaller than the \simname{GV} models.
This again indicates that the effect of reducing the magnetic wind efficiency via the $C_\beta$ parameter is as expected; the balance shifts in favor of viscous expansion and a larger disk size is obtained. 
We note that the size estimates discussed here come from the theoretical centrifugal disk and 
thus are not directly observable.
We quantitatively discuss the disk sizes and compare synthetic observations with demographics from an ALMA survey 
in Sect. \ref{subsec:obscomp}.
Additionally, we note that the gas surface density depicted in Fig. \ref{fig:global} gives us only a limited understanding of the total disk masses.
For example, the disk sizes of models \simname{GV-4}, \simname{WI-2} and \simname{WI-3a} are comparable at any given time.
However, their masses diverge significantly after the Class 0 stage. 
At one Myr, \simname{GV-4} produces a massive disk of $0.2 M_\odot$, which is nearly twice as massive as the disk in \simname{WI-2}, and over 7 times more massive than that in \simname{WI-3a}. 
The evolution of disk masses is discussed in detail in Sect. \ref{subsec:obscomp} (see Fig. \ref{fig:Mdiskevo} and Table \ref{table:final}).

\subsection{Vertical and thermo-chemical structure of the disk}
\label{subsec:synthobs}

In this section we discuss the results for the physical and thermo-chemical structure of a representative disk model obtained with the FEOSAD-{\sc ProDiMo} interface.
Figures \ref{fig:ProDiMo1} and \ref{fig:ProDiMo2} show the model \simname{WI-3a} at 0.2 and one Myr, probing the disk as it progresses through the Class II stage.
The first row of Fig.~\ref{fig:ProDiMo1} shows the evolving surface density profile $\Sigma_{\rm g}(r)$.
The early phases are featured by a relatively flat surface density between about one au and ten au, with a number of transient time-dependent pressure bumps, which are much more pronounced in dust than in gas. 
At later stages, only one pressure bump remains, whereas the disk outside of about three au follows a structure as expected from viscous spreading \citep[see, e.g.,][]{McCaughreanODell96,Hartmann+98}, with an approximate radial power-law surface density profile and an exponential decay beyond a critical radius of about 50\,au.
Inside of about one au, the surface density profile increases with the radius in both gas and dust at all evolutionary phases. 
The overall radial structure is similar to previous investigations with FEOSAD, where concentric gas and dust rings form inside of the innermost five au region of the dead zone due to viscous torques \citep{Kadam+19,Kadam+22}.  
These regions are expected to host streaming instabilities and produce planetesimals \citep{Yang+17}. 

A salient difference between an early (0.2\,Myr) and an evolved Class II disk (one Myr), according to this model, is the distribution of the dust. 
At 0.2 Myr, the global dust-to-gas mass ratio only marginally deviates from the initial value of 1\%, with the total dust and gas masses of $9\times10^{-4}\,M_\odot$ and $0.16\,M_\odot$, respectively.
In the outer disk regions, the local dust-to-gas ratio is also maintained and is only marginally less than the initial value.
However, at one Myr, most of the dust is found in the vicinity of the pressure maximum near 1 au.
Both, the total dust and gas masses, $8.6\times10^{-5}\,M_\odot$ and $0.029\,M_\odot$, respectively, have diminished during disk evolution approximately by a factor of 10.
The outer disk regions are severely depleted of dust; the local gas-to-dust ratio beyond three au is $\lesssim$\,1:10000. 
This is a consequence of the dust dynamics via its growth in the low viscosity environment and subsequent inward radial drift.

\begin{figure*}[!htbp]
\centering
\begin{tabular}{cc}
0.2\,Myr & 1.0\,Myr\\
\hspace*{-4mm}
\includegraphics[height=54mm,trim=0 49 0 0,clip]
                {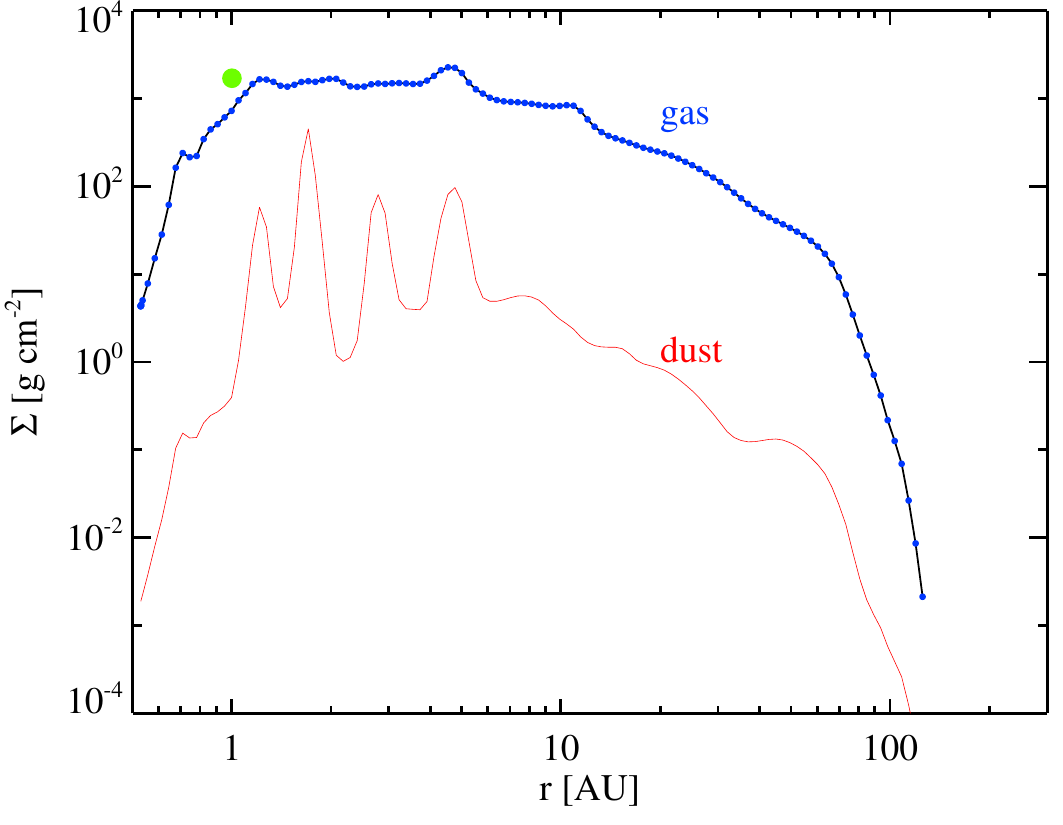} &
\hspace*{-6mm}
\includegraphics[height=54mm,trim=60 49 0 0,clip]
                {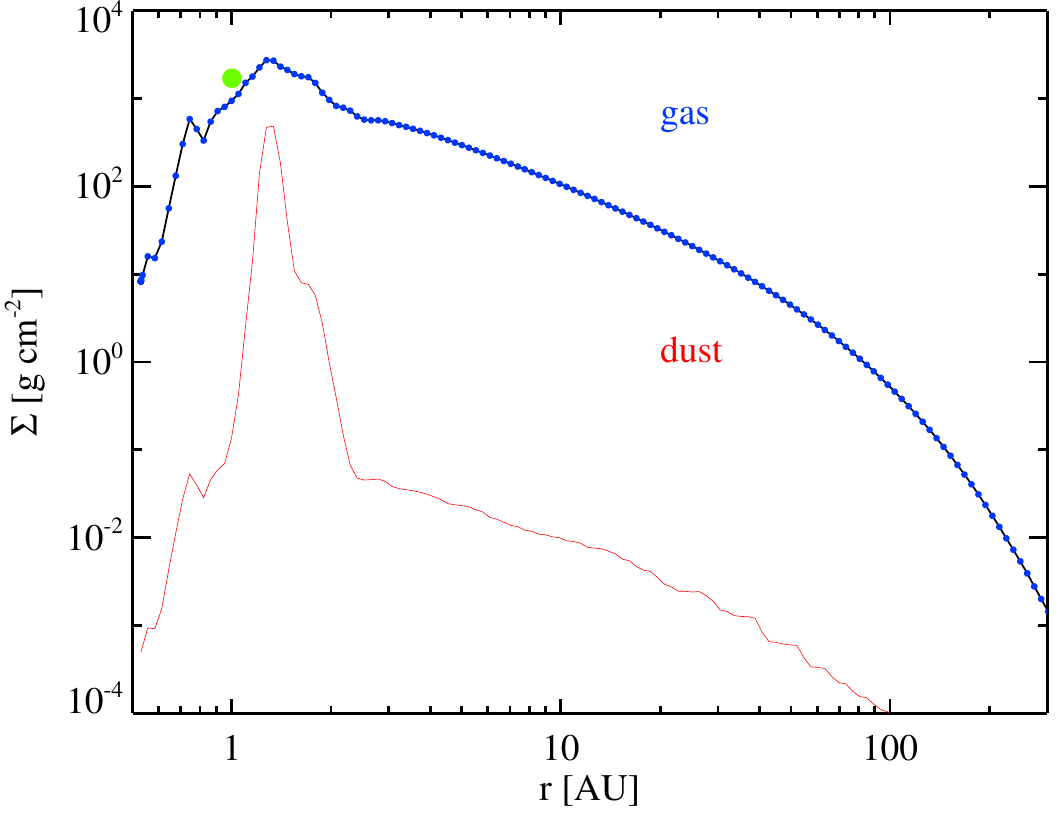} \\[-2mm]
\hspace*{-4mm}
\includegraphics[height=54mm,trim=0 49 0 0,clip]
                {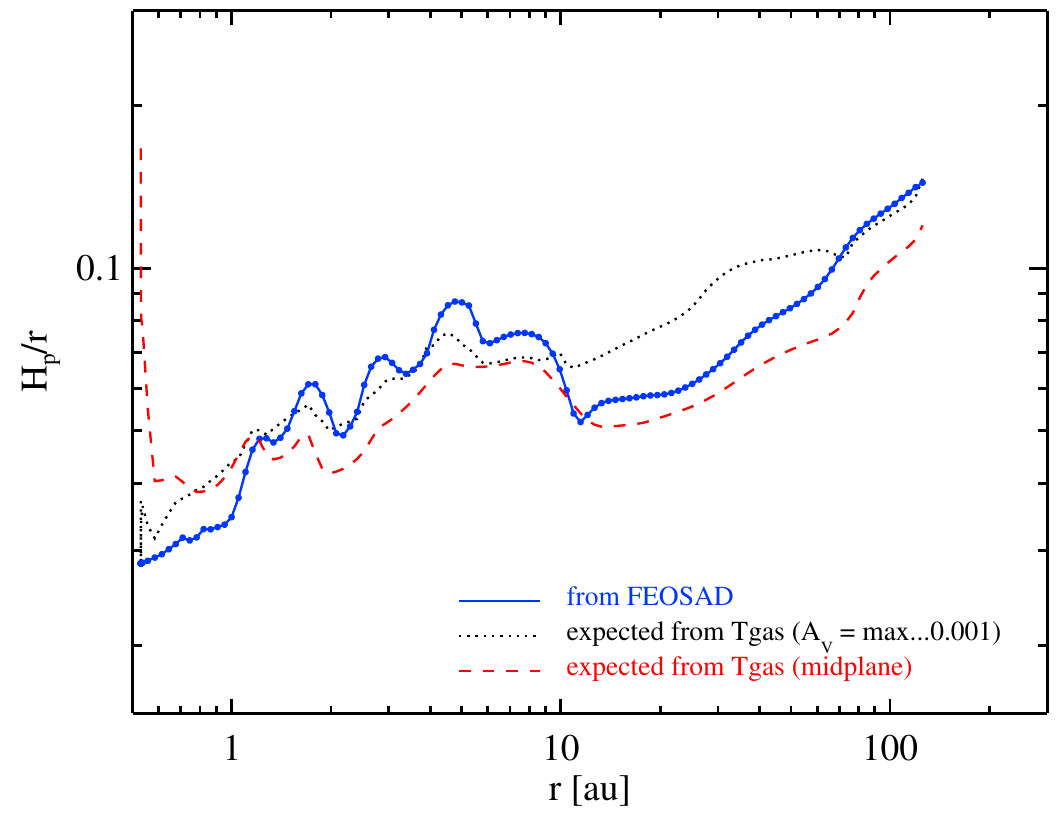} &
\hspace*{-6mm}
\includegraphics[height=54mm,trim=60 49 0 0,clip]
                {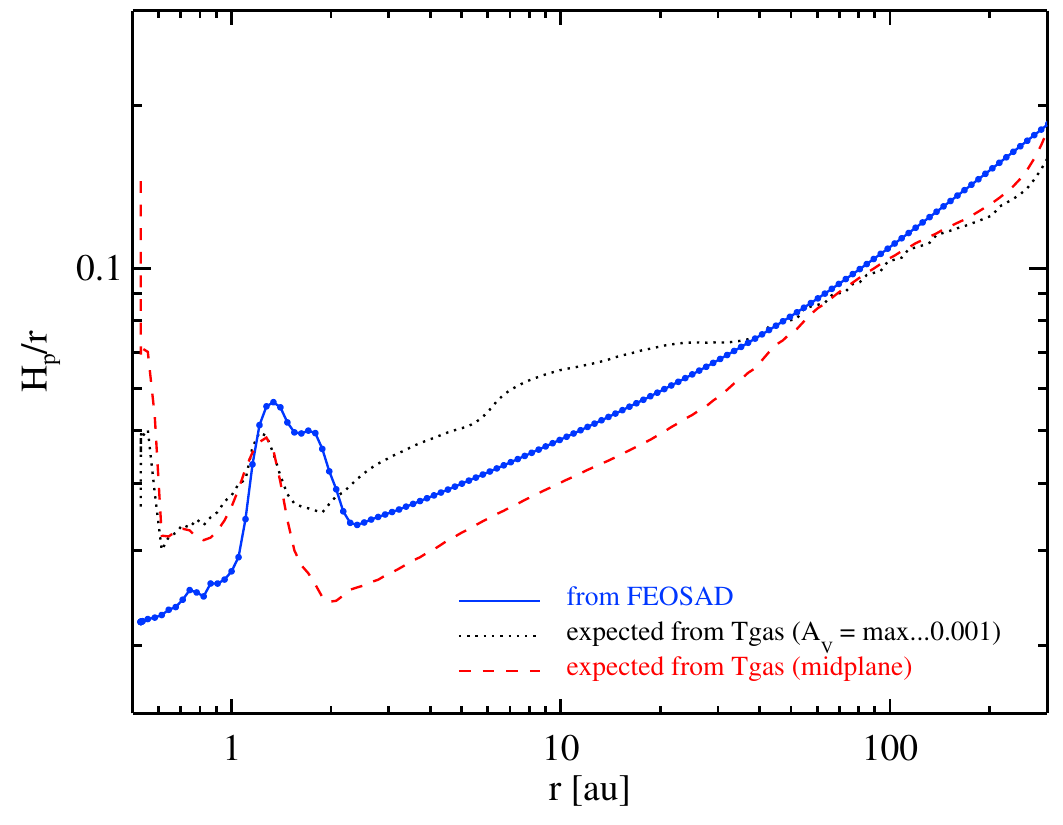} \\[-2mm]
\hspace*{-4mm}
\includegraphics[height=54mm,trim=-8 42 0 0,clip]
                {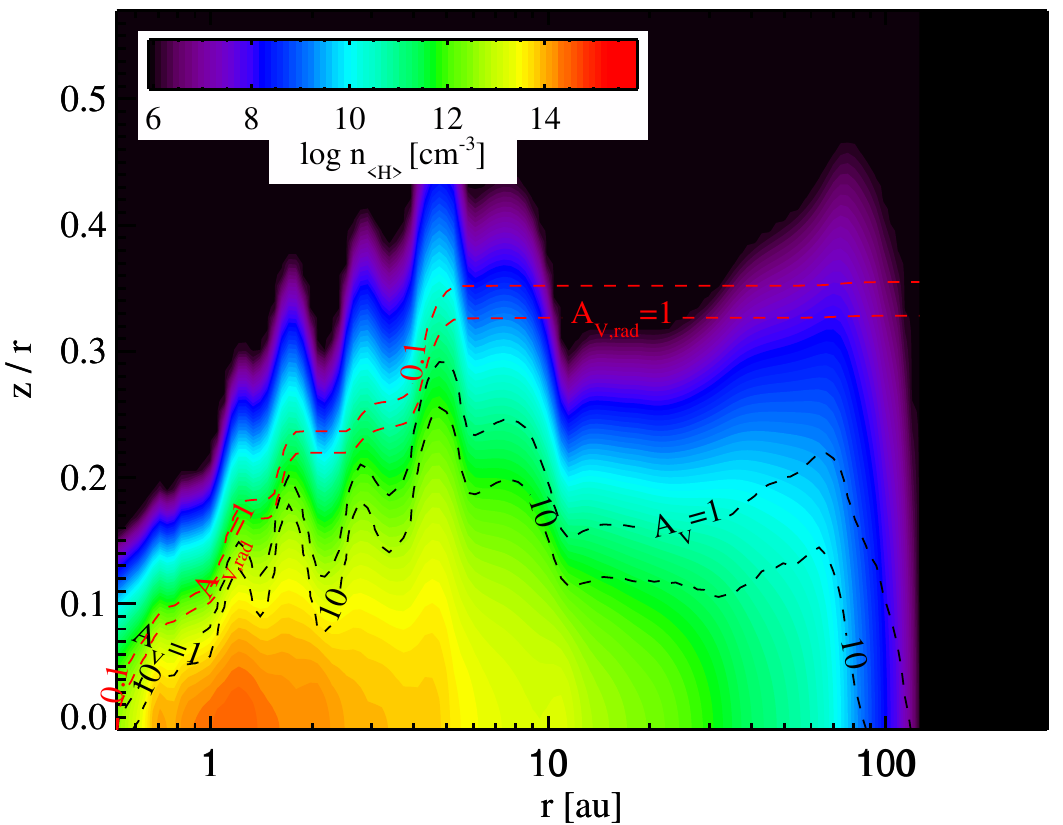} &
\hspace*{-6mm}
\includegraphics[height=54mm,trim=52 42 0 0,clip]
                {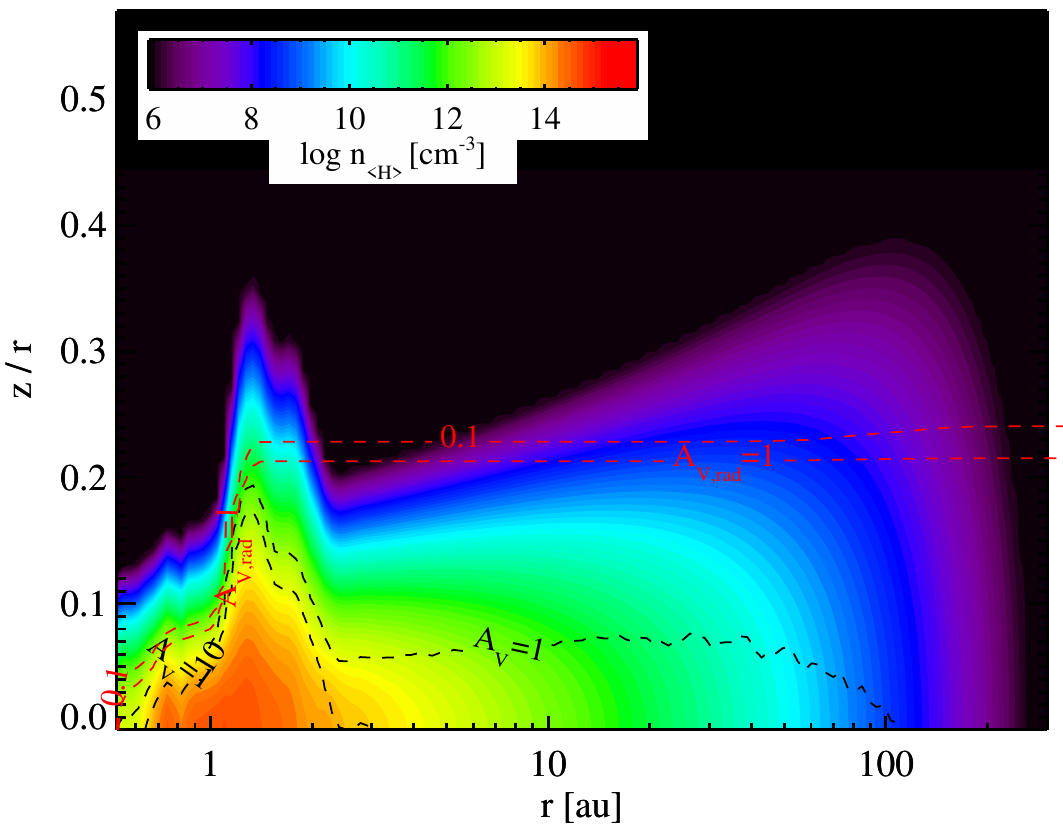} \\[-1.5mm]
\hspace*{-4mm}
\includegraphics[height=60.4mm,trim=-8 0 0 0,clip]
                {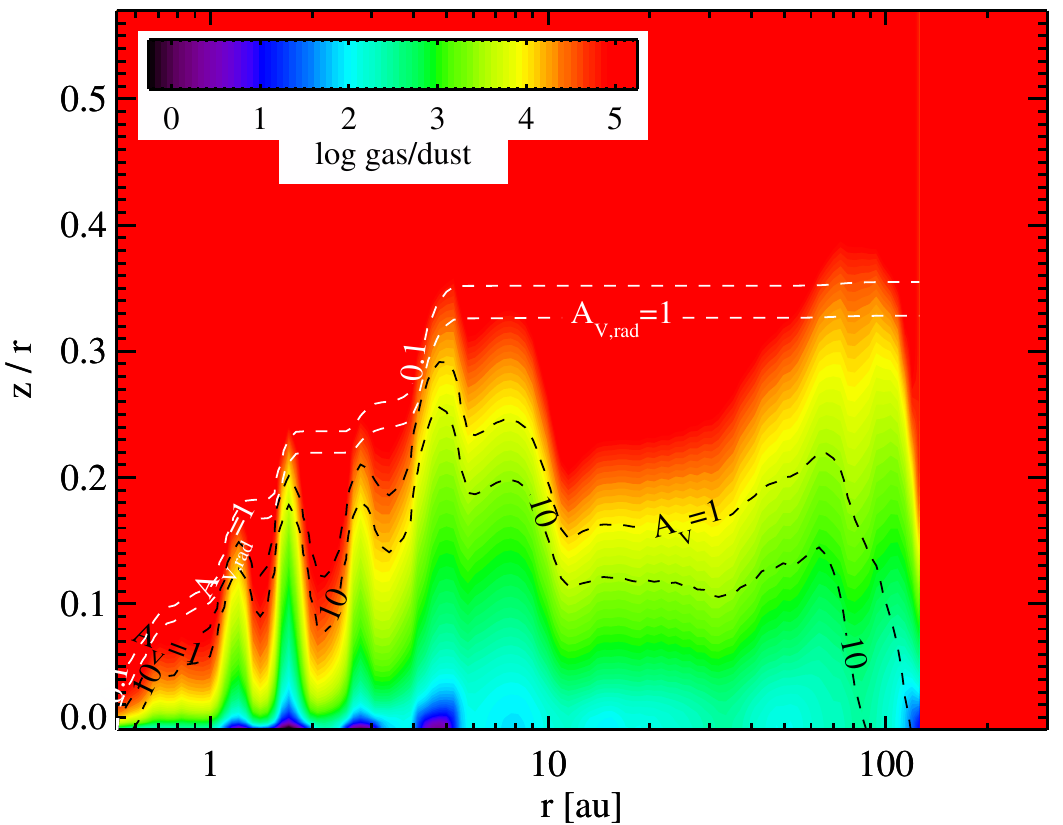} &
\hspace*{-6mm}
\includegraphics[height=60.4mm,trim=52 0 0 0,clip]
                {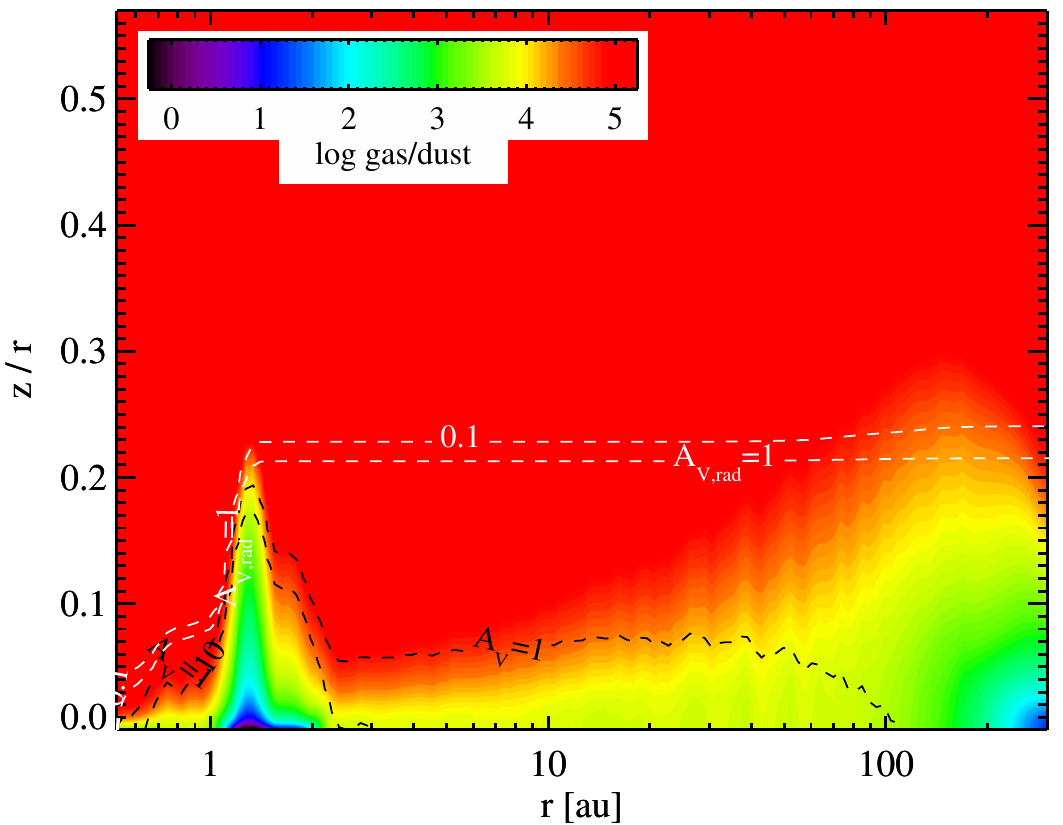} \\[-2mm]
\end{tabular}
\caption{Disk structure of the \simname{WI-3a} model after 0.2\,Myr and one Myr.  The upper plots show the gas and dust column densities, where the green dot marks the popular MMSN-value of $\rm 1700\,g/cm^2$ at one au. The second row of plots shows the scale heights $H_g(r)$ as passed from FEOSAD, and the values calculated from the gas temperatures calculated by {\sc ProDiMo}, using either the midplane temperature only (red dashed) or the $T_{\rm gas}(z)$ structure up to a height where the radial visual extinction is $A_{\rm V,rad}\!=\!0.001$. The lower two rows show the hydrogen nuclei density $n_{\langle H\rangle}$ and the gas-to-dust mass ratio. Additional contour lines show the radial and vertical optical extinctions, $A_{\rm V,rad}$ and $A_{\rm V,ver}$, respectively.}
\label{fig:ProDiMo1}
\end{figure*}

\begin{figure*}
\centering
\vspace*{-3mm}
\resizebox{!}{115mm}{  
\begin{tabular}{cc}
0.2\,Myr & 1.0\,Myr\\
\hspace*{-4mm}
\includegraphics[height=54mm,trim=-10 42 0 0,clip]
                {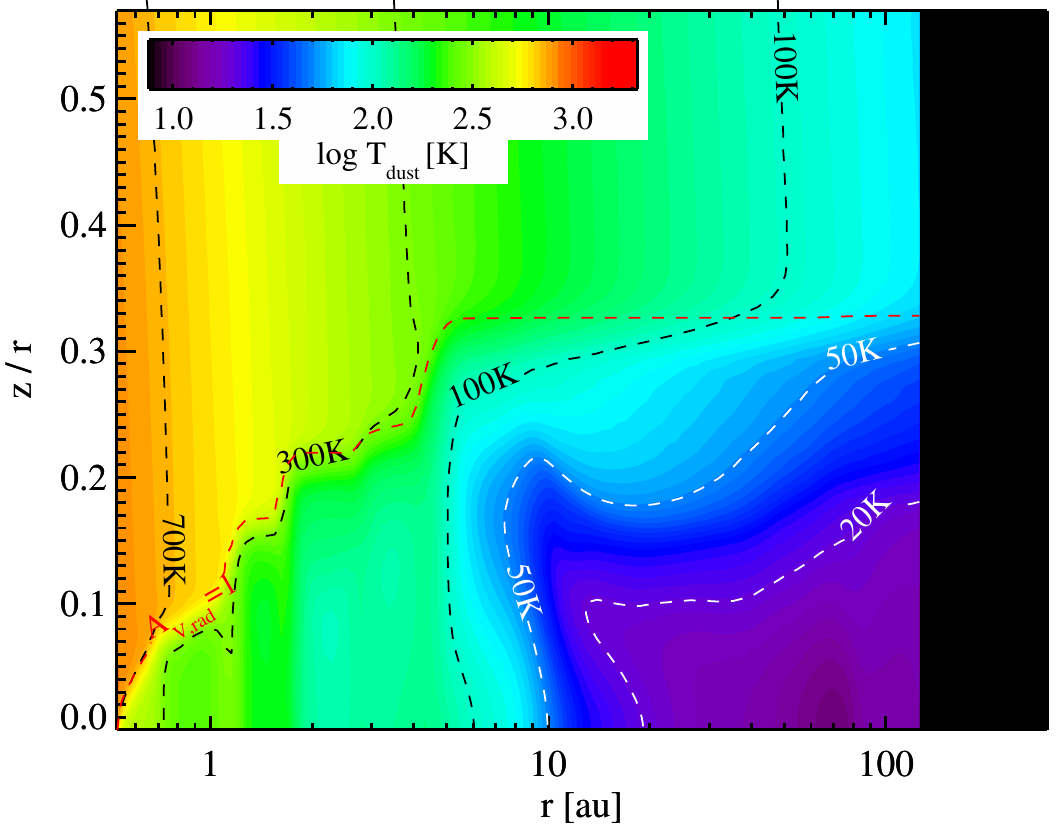} &
\hspace*{-6mm}
\includegraphics[height=54mm,trim=50 42 0 0,clip]
                {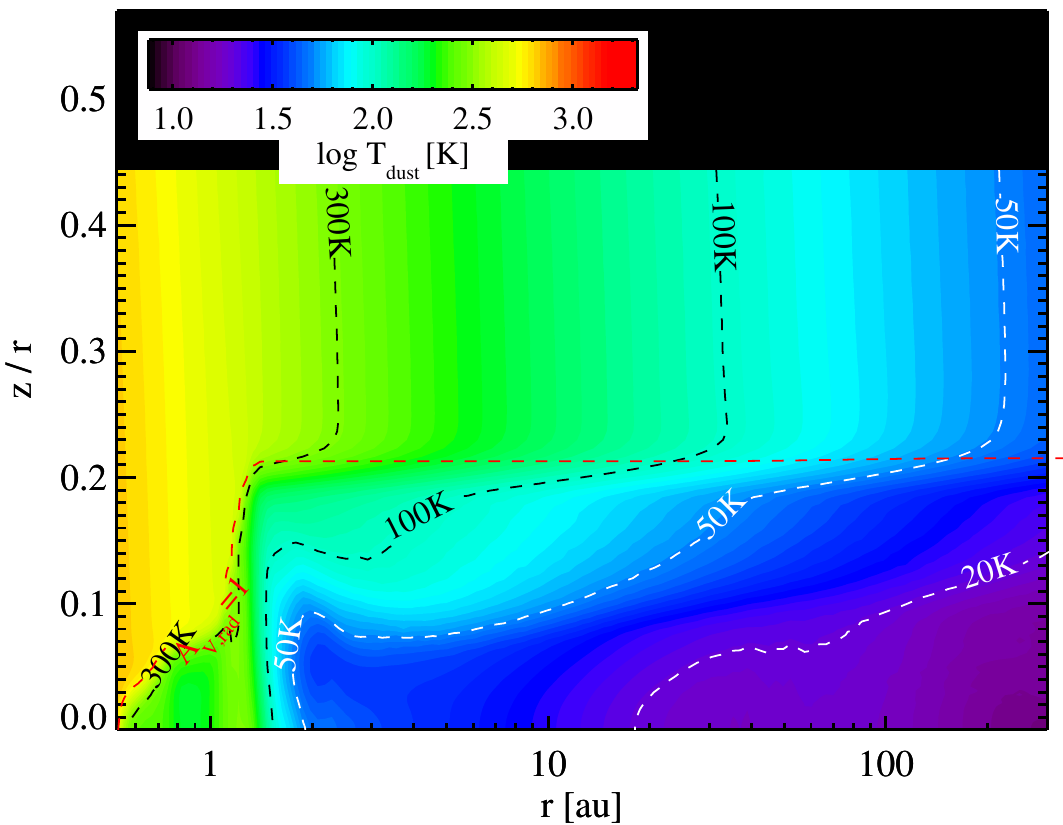} \\[-1mm]
\hspace*{-4mm}
\includegraphics[height=60.4mm,trim=-10 0 0 0,clip]
                {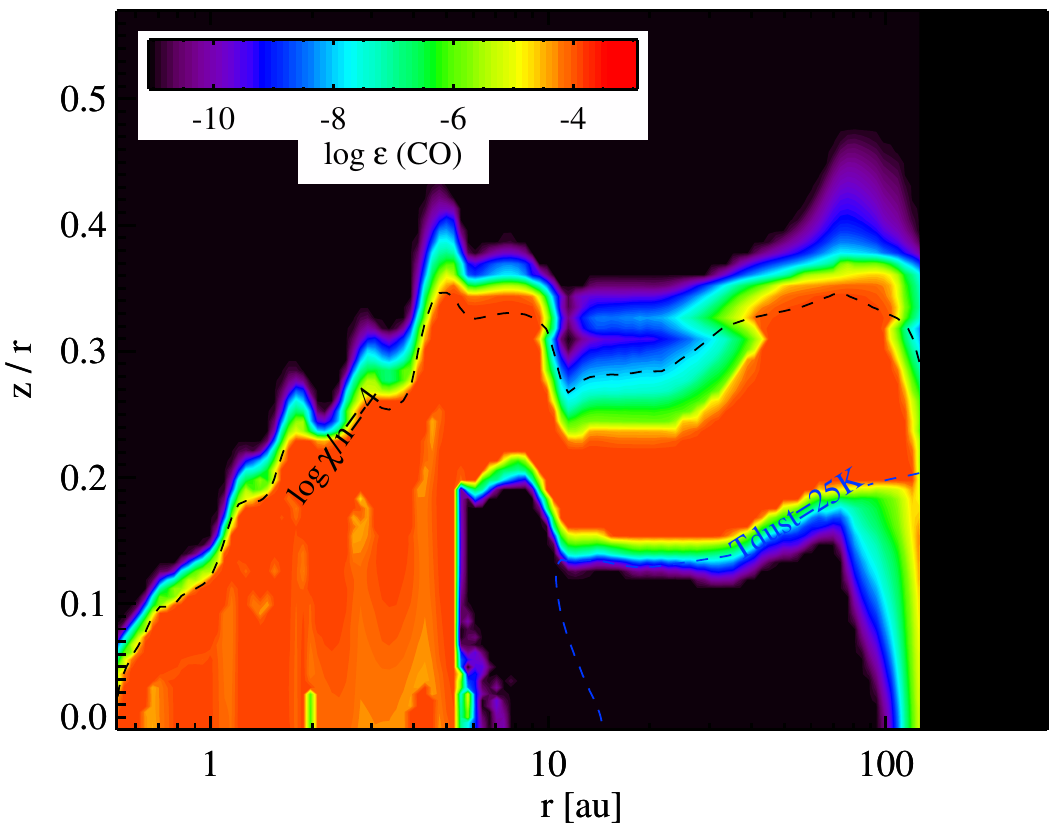} &
\hspace*{-6mm}
\includegraphics[height=60.4mm,trim=50 0 0 0,clip]
                {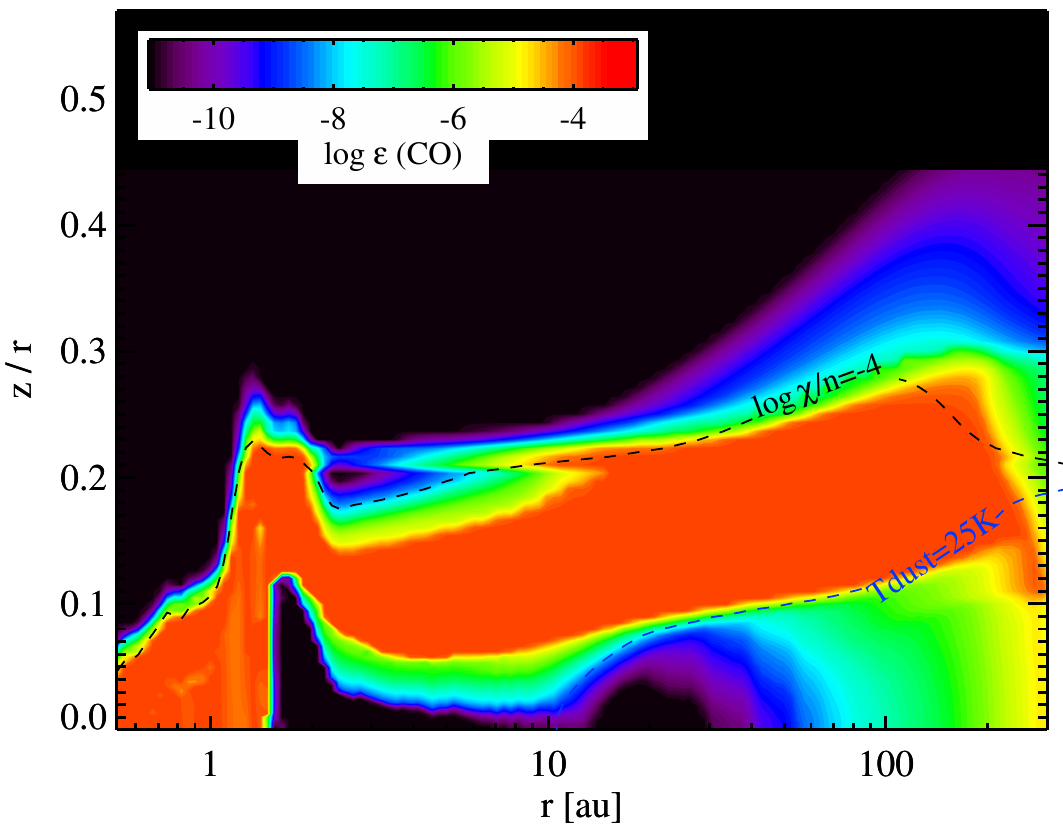} \\[-6mm]
\hspace*{-6mm}
\includegraphics[height=65mm,trim=30 50 60 350,clip]
                {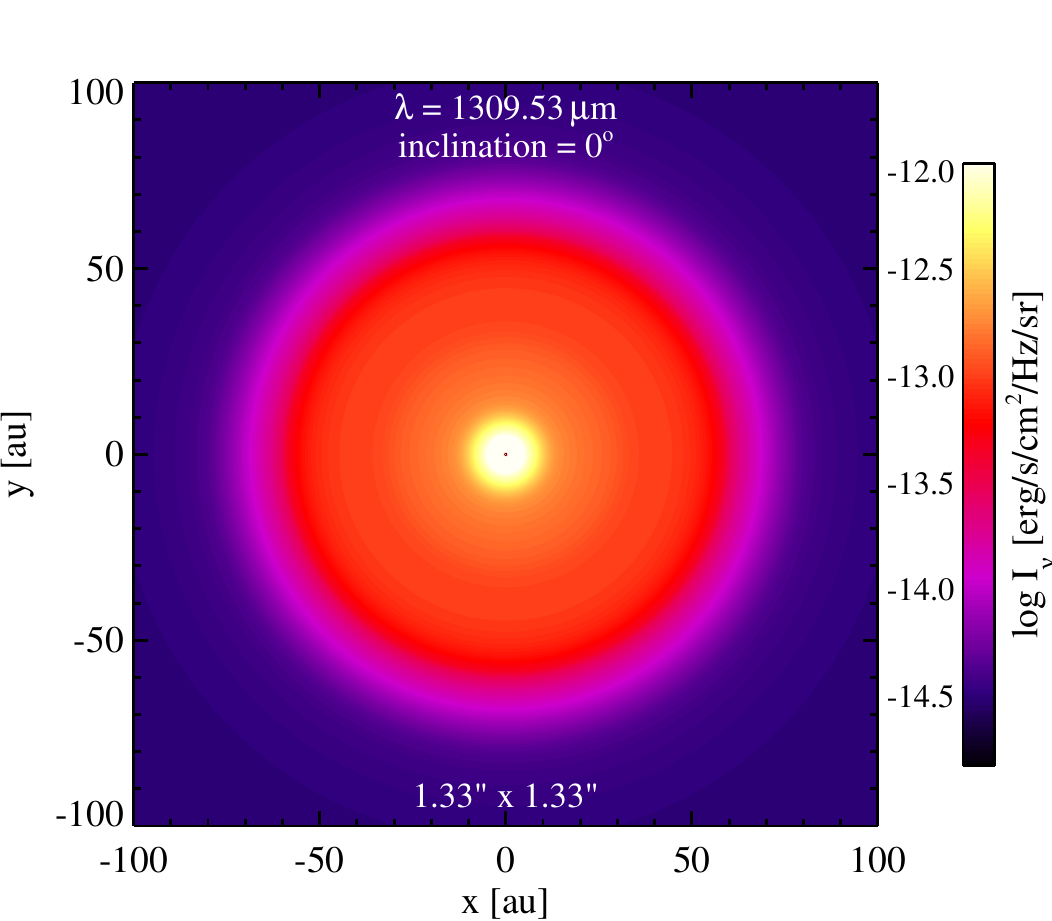} &
\hspace*{-6mm}
\includegraphics[height=65mm,trim=65 50 60 350,clip]
                {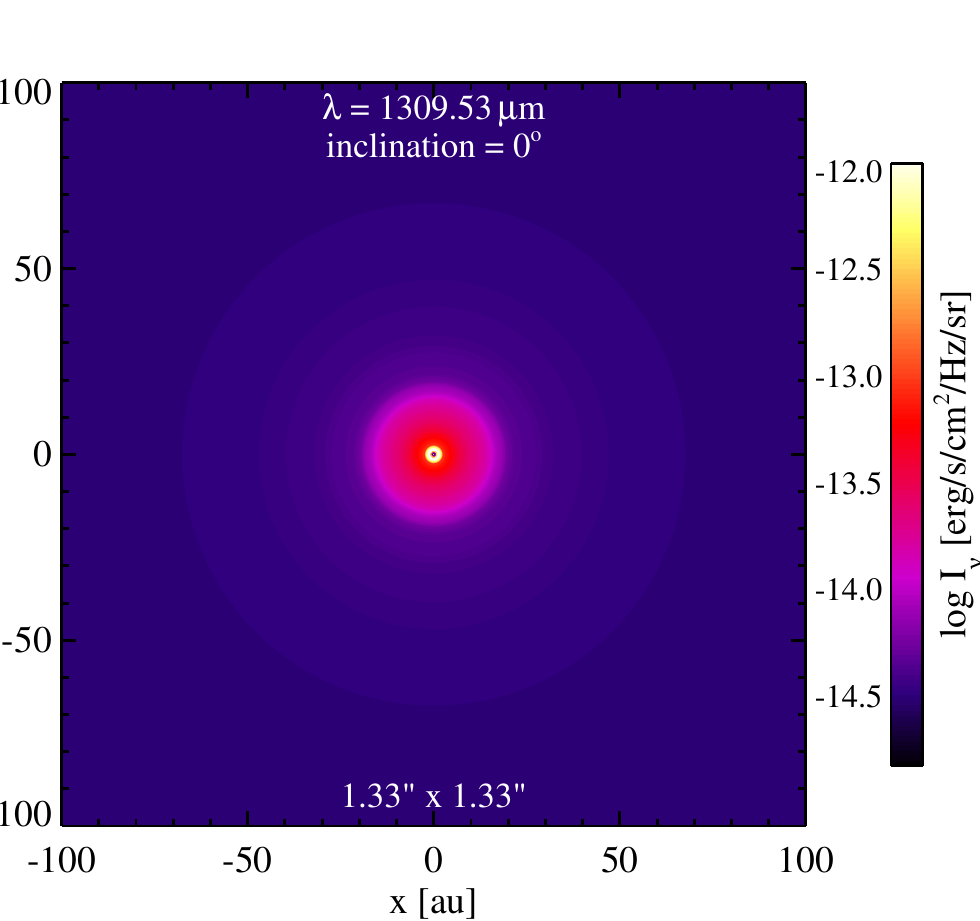} \\[-1mm]
\hspace*{-15mm}
\includegraphics[height=64mm,trim=30 31 60 350,clip]
                {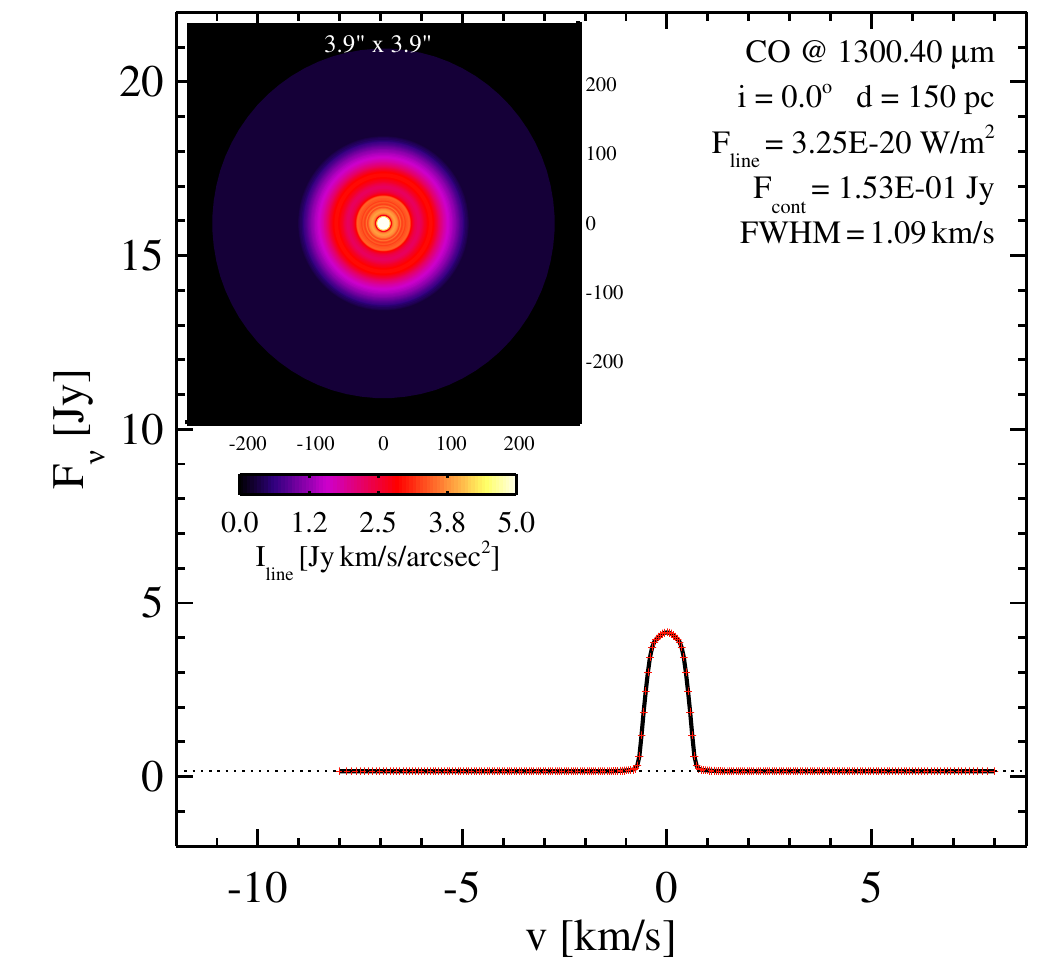} &
\hspace*{-20mm}
\includegraphics[height=64mm,trim=30 31 60 350,clip]
                {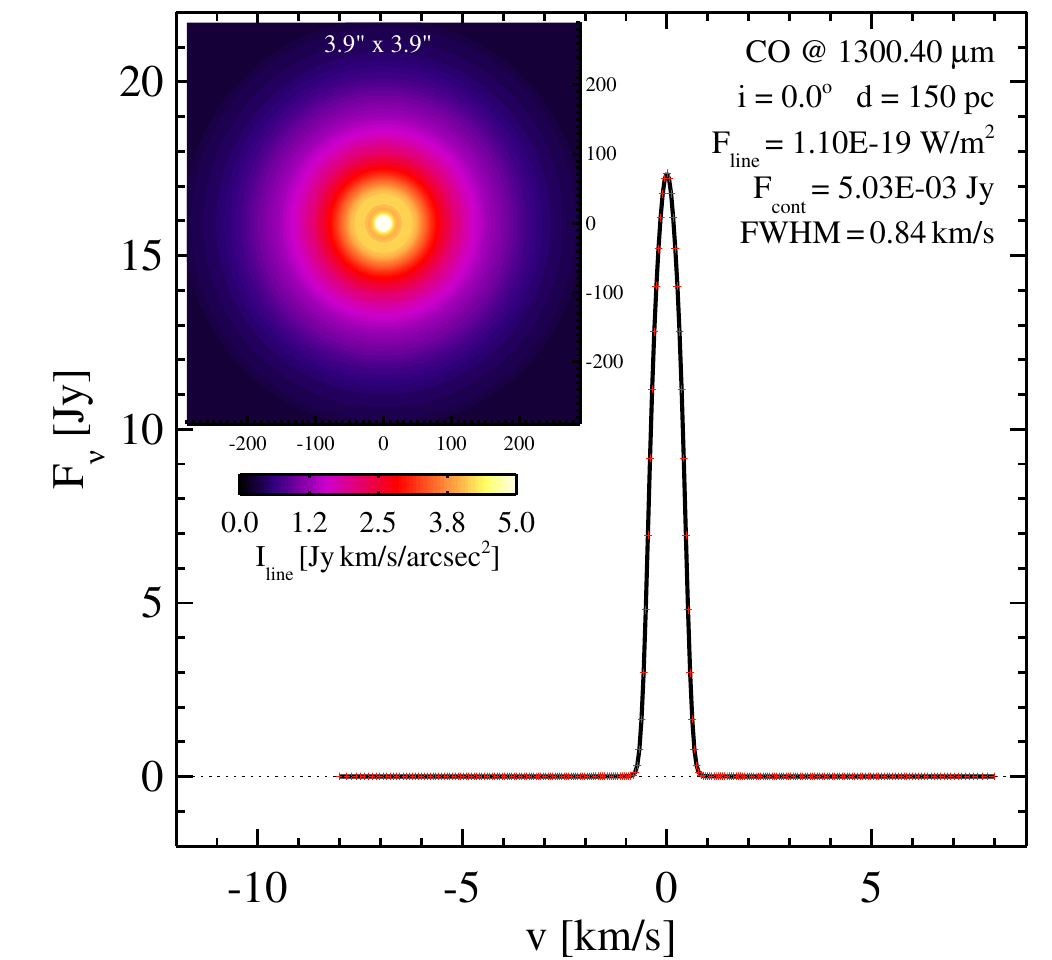} \\[-1mm]
\end{tabular}}
\caption{Continuation of Fig.~\ref{fig:ProDiMo1} showing the dust temperature structure (top) and the CO-concentration (second row) calculated by {\sc ProDiMo} based on the \simname{WI-3a} model. Additional contour times show the ionization parameter $\chi/n$ (UV field strength divided by total particle density) to indicate where about the CO molecules photo-dissociate, and a dust temperature of 25\,K to indicate where CO freezes out. The lower diagrams show the simulated 1.3 continuum and CO 2-1 line maps, from which we derive a continuum radius of 89\,au and 109\,au at 1.3\,mm after 0.2\,Myr and one Myr, respectively, and a disk radius in the CO 2-1 line of 198\,au and 375\,au, respectively. Each radius encircles 90\% of the respective continuum or line flux.}
\label{fig:ProDiMo2}
\end{figure*}

In the second row of Fig.\,\ref{fig:ProDiMo1}, we compare the scale heights calculated from FEOSAD's midplane temperature and taking disk self-gravity into account (blue), to that calculated from {\sc ProDiMo}'s midplane temperature (red dashed), and a more sophisticated variant thereof (black dotted).
In case of the latter, the equation of hydrostatic equilibrium is integrated upward into the optically thin region using {\sc ProDiMo}'s $T_{\rm gas}(r,z)$ and then fitted by a Gaussian of single temperature.
In the early phase, the disk is taller in the vicinity of inner rings, because of the combined effect of viscous heating and large vertical dust optical depths, which hinder the viscous heat to escape and create a temperature inversion (Fig. \ref{fig:ProDiMo2}). 
As the disk evolves, the viscous heating diminishes and the scale heights in the inner regions decrease.
Hence, the disk becomes increasingly more extended and flared due to the impact of direct irradiation from the star.
Since FEOSAD and {\sc ProDiMo} use different opacities (see Sects.\,\ref{subsec:disk} and \ref{subsec:prodimo}), it is no surprise that the independent computation of the scale heights by {\sc ProDiMo} is not identical with the values passed from FEOSAD.
However, the overall magnitude and shape of the calculated $H_{\!g}(r)$ matches reasonably well, especially in the outer disk beyond 10\,au.
Since the inner rim near the sink cell is directly illuminated by the star in {\sc ProDiMo}, it is much warmer and the scale height is much larger as compared to that in FEOSAD, where this radiative effect is not included.

The gas density structure shown in the third row of Fig. \ref{fig:ProDiMo1} directly follows from $\Sigma_{\rm g}(r)$ and $H_{\!g}(r)$.
We obtain maximum gas particle densities of about $10^{15}\rm\,cm^{-3}$ in the pressure bumps around one au. 
The last row of Fig. \ref{fig:ProDiMo1} shows the gas-to-dust mass ratio after settling.  
We obtain local dust-to-gas ratios $\ga\!1$ after dust settling in the pressure bumps, which is consistent with the midplane values in FEOSAD.  
However, the dust-to-gas ratio is $<$\,10$^{-4}$ at the disk surface where the optical radial extinction $A_{\rm V,rad}\!=\!1$, which is an important result concerning the formation of mid-IR molecular emission lines observable with Spitzer and the James Webb Space Telescope (JWST) \citep[][]{Woitke2023}.

In Fig.\,\ref{fig:ProDiMo2}, the first row shows the dust temperature structure as calculated by {\sc ProDiMo}. 
Due to viscous heating and close-to-diffusive radiative transfer, the midplane temperature features several smaller variations between one and five au at an age of 0.2\,Myr. 
Consequently, as seen in the isothermal contours, there are multiple icelines (or snow-surfaces in 2D) around 150\,K behind the pressure bumps, beyond five au. 
At the later stage, the icelines sit right behind the first and only pressure bump at about 1.5\,au.
{Considering the second row of Fig.\,\ref{fig:ProDiMo2}, the CO molecules populate a sandwich layer between disk midplane and its surface in the outer disk beyond the pressure bumps.} 
This narrow layer is (i) warm enough to prevent CO freeze-out ($T_{\rm d}\!\ga\!25\,$K), and (ii) protected enough from UV radiation to prevent CO photo-dissociation. 
The latter indicated by $\chi/n\!\la\!10^{-4}$ line, where $\chi$ is the UV field strength normalize to standard solar neighborhood, and $n$ is the total hydrogen nuclei density.

\begin{table*}[!t]
  \centering
  \caption{Simulated ALMA band~6 and band~7 continuum and CO line fluxes at a distance of 150\,pc and disk radii for the \simname{WI-3a} model, predicted by {\sc ProDiMo}, using different tracers and threshold flux levels.  
  }
  \label{tab:Rdisk}
  \vspace*{-2mm}
  \begin{tabular}{r|ccccc|ccccccc}
    \hline
    &&&&&\\[-2.1ex]
    & \multicolumn{5}{c|}{model \simname{WI-3a} at 0.2\,Myr}
    & \multicolumn{5}{c}{model \simname{WI-3a} at 1.0\,Myr} \\
    &&&&&\\[-2.1ex]
                        & flux     &  80\%  &  90\%  &  95\%   &  98\%
                        & flux     &  80\%  &  90\%  &  95\%   &  98\%\\
    \hline
    &&&&&\\[-2.1ex]
    continuum 1.3\,mm   & 150\,mJy &  49\,au & {  55\,au} &  60\,au & 67\,au
                        &   5\,mJy &  19\,au & {  51\,au} &  97\,au & 170\,au\\
  continuum 870\,$\mu$m & 350\,mJy &  53\,au & {  59\,au} &  64\,au & 71\,au
                        &  15\,mJy &  26\,au & {  60\,au} & 107\,au & 179\,au\\
    \hline
    &&&&&\\[-2.1ex]
          CO 2-1 &  4.2\,Jy\,km/s &  95\,au & { 106\,au} & 113\,au & 119\,au
                 & 14.3\,Jy\,km/s & 227\,au & { 268\,au} & 292\,au & 314\,au\\
    Band 6 \,\,\,   $^{13}$CO 2-1 & 1.7\,Jy\,km/s &  86\,au & { 97\,au} & 103\,au & 109\,au
                 &  4.2\,Jy\,km/s & 163\,au & { 193\,au} & 219\,au & 250\,au\\
    C$^{18}$O 2-1 & 0.9\,Jy\,km/s &  79\,au & {  88\,au} &  95\,au & 102\,au
                 &  1.8\,Jy\,km/s & 131\,au & { 160\,au} & 185\,au & 216\,au\\
    \hline
    &&&&&\\[-2.1ex]
    CO 3-2 &   9.7\,Jy\,km/s &  95\,au & { 106\,au} & 113\,au & 119\,au
                 &  32.2\,Jy\,km/s & 226\,au & {269\,au} & 294\,au & 317\,au\\
    Band 7 \,\,\,   $^{13}$CO 3-2 &  4.2\,Jy\,km/s &  87\,au & { 97\,au} & 103\,au & 110\,au
                  & 10.4\,Jy\,km/s & 162\,au & { 192\,au} & 217\,au & 248\,au\\
    C$^{18}$O 3-2 &  2.3\,Jy\,km/s &  81\,au & { 90\,au} &  96\,au & 103\,au
                  &  4.6\,Jy\,km/s & 134\,au & { 161\,au} & 185\,au & 213\,au\\
    \hline
  \end{tabular}
\end{table*}  

The simulated continuum and line images are shown in the lower half of Fig.~\ref{fig:ProDiMo2}.
For deriving synthetic observations of the disk sizes in dust ($R_{\rm dust, \, Band 6}$) and gas ($R_{\rm gas,\,(CO \,2-1)}$), we consider the radius that contains 90\% of the spectral flux, which is applied to the simulated continuum images and the CO line maps.
The CO gas radii found this way are always significantly larger than the continuum radii (see Table~\ref{tab:Rdisk}), because of two effects.  
Firstly, the CO lines are an optically thick tracer of the gas, whereas, excluding the innermost regions with dust rings, the 1.3\,mm continuum (ALMA Band 6) is an optically thin tracer for the dust.
The dust signal vanishes proportional to $\Sigma_{\rm d}$. However, the CO signal 
features a slower decline with radius, reflecting the gas temperature radial profile, which is shallower than that of $\Sigma_{\rm g}$,
until it vanishes quickly as the line becomes optically thin. This first effect is always at work and it produces larger gas radii as compared to dust radii. 
Secondly, at the later stages of the disk evolution, the continuum disk radii are also affected by the dust growth followed by inward radial drift in the outer disk regions. 
This effect is evident from the much lower dust-to-gas ratios $ \lesssim 10^{-4}$ after one Myr in the outer disk (see the first row of Fig.~\ref{fig:ProDiMo1}).  In addition, those few dust grains remaining in the outer disk are the small grains, which have only little millimeter opacity.  According to the FEOSAD model, the grains at 50\,au have a maximum size of 500\,$\mu$m after 0.2\,Myr, but only 16\,$\mu$m after one Myr.
At late evolutionary stages, this second effect dominates, leading to an apparently shrinking dust radius while the gas radius is still expanding. 

Table~\ref{tab:Rdisk} shows how the derived continuum and CO line radii depend on the applied cut-off flux level, and on the choice of the line and ALMA band.
The results are found to be rather independent of the choice of the ALMA band.
However, the choice of flux threshold level is important, in particular for the continuum, which implies a considerable measurement uncertainty for the continuum radii. 
According to our model \simname{WI-3a}, different threshold flux levels result in rather similar dust radii 50-70\,au after 0.2\,Myr, but in a large variety of values between 20 and 180\,au after one Myr. 
With time, the dust grows and drifts, moving the carriers of the millimeter opacity inward. 
Despite the action of the winds, the outer disk still experiences gravitoviscous spread, which contains trace abundances of larger grains.
This explains the spread in radii with threshold flux at later times and implies that observations with high signal-to-noise ratio may result in much larger continuum radii.  
In summary, from our reference wind model \simname{WI-3a}, we conclude that the disks tend to become fainter in the continuum with age, whereas they become brighter and larger in the CO lines due to gravitoviscous spreading.

\subsection{Magnetic disk winds and observational constraints}
\label{subsec:obscomp}

The disk size and the mass are some of the most fundamental characteristics of a protoplanetary disk and some of the observable population-level differences between disks evolving with and without magnetic winds manifest in these global properties.  
In this section we compare our magnetic disk wind models against those evolving only with gravitoviscous forces, in the context of observational constraints.
Here we briefly summarize the complications of observing protoplanetary disks (see \citet{Miotello+23} for a recent review). 
When measuring disk size, the dust radius is typically measured from surface brightness profile of dust continuum emission in millimeter wavelengths; the most favored for ALMA surveys is Band 6 at 1.3 mm.
The radius is defined empirically as the radius which contains a certain fraction of the total disk emission (68\%, 90\%, or 95\%).
The gas emissions from a protoplanetary disk can be measured via molecular line traces such as CO isotopologs. 
For measuring the disk size in gas, the zero moment map of $^{12}$CO (J=2–1) is typically utilized, which is optically thick and easier to detect in small amounts at large radii.
The gas radius is calculated similar to the methods as described above and it is observed to be larger than the dust radius, typically by a factor of a few \citep{Ansdell+18}. 
As discussed in Sect. \ref{subsec:synthobs}, the larger gas disk radii can be attributed to optical depth effects as well as dust drift.

With the assumptions of optically thin emission along with a prescription of opacity and disk temperature, the continuum flux can be translated into the dust mass \citep{Hildebrand83}, although these assumptions do not always necessarily hold \citep[][]{Dunham+14,Woitke+19,Haworth21}. 
{Direct measurements of the gas mass, which is expected to contain the majority of the disk mass, are difficult due to the faintness and inaccessibility of H$_2$ line observations. 
However, the method of measuring gas masses using a tracer such as CO lines has severe limitations arising from the uncertainty in the CO/H ratio in disks {(more specifically, gas-phase [CO]/[H$_2$] ratio, see Fig. \ref{fig:ProDiMo2})}, which is inevitably connected to the disk thermochemistry \citep[][]{Ansdell+16,Bosman+21}. 
Traditionally, the dust mass measured from continuum flux  is translated into the total disk mass through a multiplicative factor of 100, as indicated by the interstellar gas-to-dust ratio.
Predictions from hydrodynamic simulations of PPDs, however, point toward significant local variations from this value due to dust dynamics \citep{Takeuchi-Lin02,Vorobyov-Elbakyan19} 
Thus, these different approaches of measuring the mass do not agree with each other for PPDs around low mass stars, resulting in significant uncertainties.}

With these caveats in mind, we compare our results with the latest compilation of ALMA survey data of Class II protoplanetary disks.
The star forming regions considered here are Ophiuchus, Taurus, Lupus and Chamaeleon, which have similar, relatively young ($\sim 0.8-3$ Myr) ages and form a near complete sample \citep{Ansdell+16, Ansdell+18,Manara+23,Miotello+23}.
Since not all data is available for each of the star forming regions, Table \ref{table:sfregions} summarizes the data used for comparison with the simulated models. 
{The theoretical masses and radii resulting from simulations are not directly observable, which motivates us to make synthetic observations with {\sc ProDiMo} as elaborated in Sect. \ref{subsec:synthobs}. 
Here we compare the synthetic observations in terms of radii $R_{\rm{dust, Band 6}}$ and $R_{\rm gas, ^{12}{CO}}$, both calculated at 90\% flux threshold.}
The dust content of the disk $M_{\rm dust, Band 6}$ is also compared with the survey data, while for the gas component, we use the line fluxes of ${\rm ^{12}{CO}}$ and ${\rm {C^{13}O}}$ at a standard distance of 150 pc.

\begin{figure}
\centering
  \includegraphics[width=0.45\textwidth]{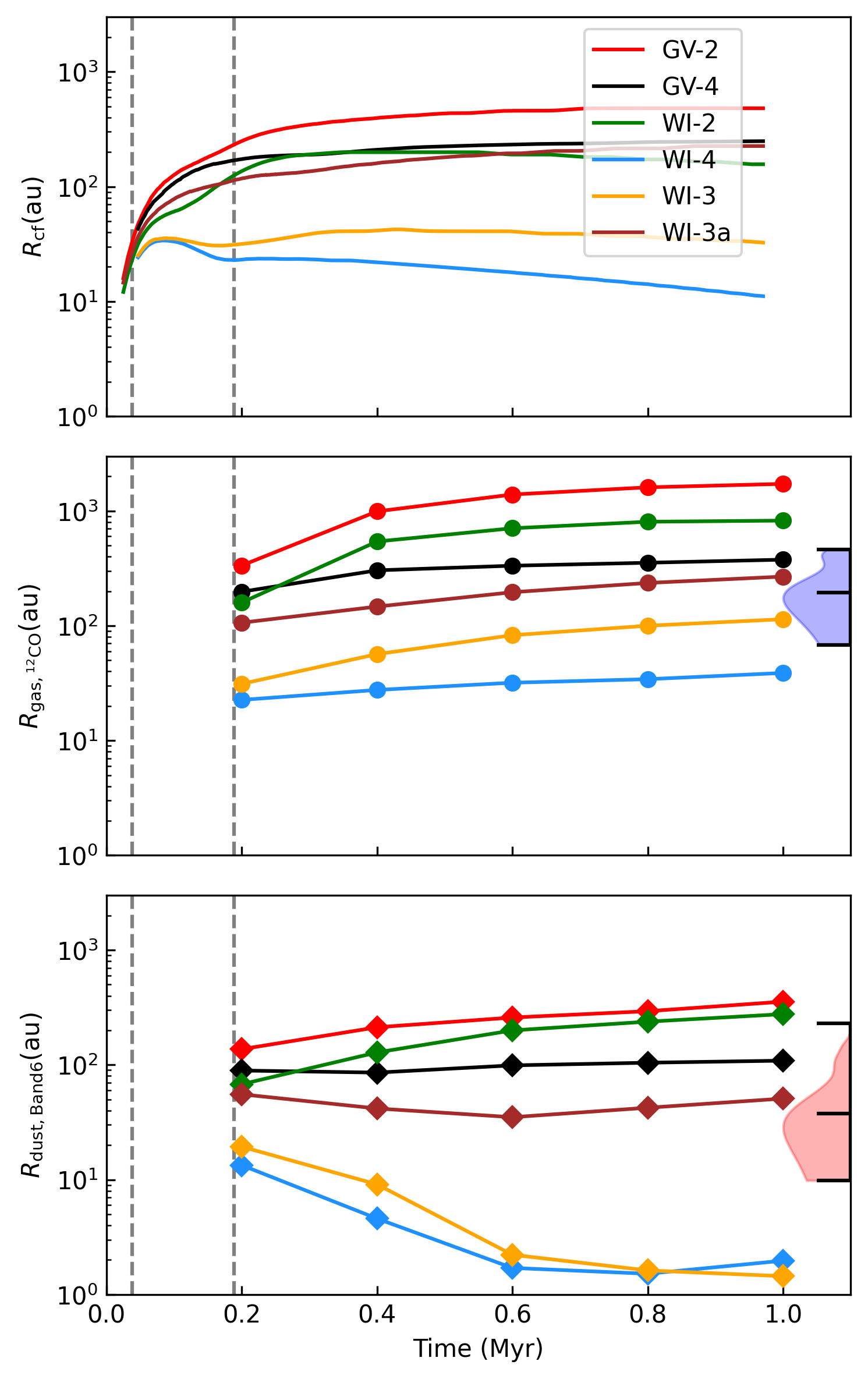}
\caption{Evolution of the disk size. Top panel: Theoretical centrifugal disk radii. Bottom two panels: Radii obtained from {\sc ProDiMo} synthetic observations . 
The gas and dust radii for the models are calculated via synthetic $^{12}$CO and Band 6 continuum emissions with 90\% flux threshold.
The vertical, dashed gray lines mark the average Class 0/I and Class I/II boundaries. The shaded regions on the right depict half-violin plots of gas (red) and dust (red) radii of the Class II disks, as observed in a subset of ALMA surveys (see Table \ref{table:sfregions}).}
\label{fig:Rdiskevo}
\end{figure}

In Fig. \ref{fig:Rdiskevo} we show the temporal evolution of the centrifugal radius as well as synthetic gas and dust radii for the models.
The vertical dotted lines in these plots indicate the average location of the Class 0/I and Class I/II boundaries, as defined in Sect. \ref{subsec:global}.
The top panel of this figure shows the theoretical size evolution of the disks in terms of $R_{\rm cf}$, which  
is smoothed with a box-car average with a box-width of 0.05 Myr, in order to get rid of stochastic variations caused by disk size pulsation inherit to young stellar objects with variable protostellar accretion \citep{Vorobyov+20}.   
The significantly smaller sizes of low viscosity disks with winds in \simname{WI-3} and \simname{WI-4} are apparent in these plots.
Even in the early Class 0 and I stages, these disks are much smaller than their gravitoviscous counterparts. 
As also seen in Fig. \ref{fig:global}, the disks in models \simname{GV-4}, \simname{WI-2} and \simname{WI-3a} cluster around 150-250 au, while \simname{GV-2} produces a large viscously spreading disk.
The exact values of disk radii as well as some of the global disk properties at one Myr are listed in Table \ref{table:final} for comparison.

The second and third panels of Fig. \ref{fig:Rdiskevo} show synthetic measurements of gas and dust radii, respectively.
As elaborated in Sect. \ref{subsec:synthobs}, the gas radius is approximated from $^{12}$CO(J=2-1) line emissions, {while the dust radius is estimated from simulated 1.3 mm continuum observations, each with a 90\% flux threshold. }
The synthetic observations obtained via {\sc ProDiMo} are performed after the embedded Class 0 and Class I stages are over, when the disk may be visible through the nascent cloud code.
Starting at 0.2 Myr, a total of five {\sc ProDiMo} calculations are performed per model, which are spaced uniformly in time so that the observational trends can be followed as a PPD evolves through the Class II stage. 
The shaded regions near the right axis of these panels depict half-violin plots of observed gas and dust radius data.
The gas radii pertain to Lupus cluster only \citep{Ansdell+18}, while the dust radii are obtained from all four aforementioned star forming regions \citep{Hendler+20,Manara+23}.
{These violin plots show the probability distribution functions for the data,} which are terminated at the extreme values denoted by dashes, with the central dash showing the median value.

\begin{figure}
\centering
  \includegraphics[width=0.47\textwidth]{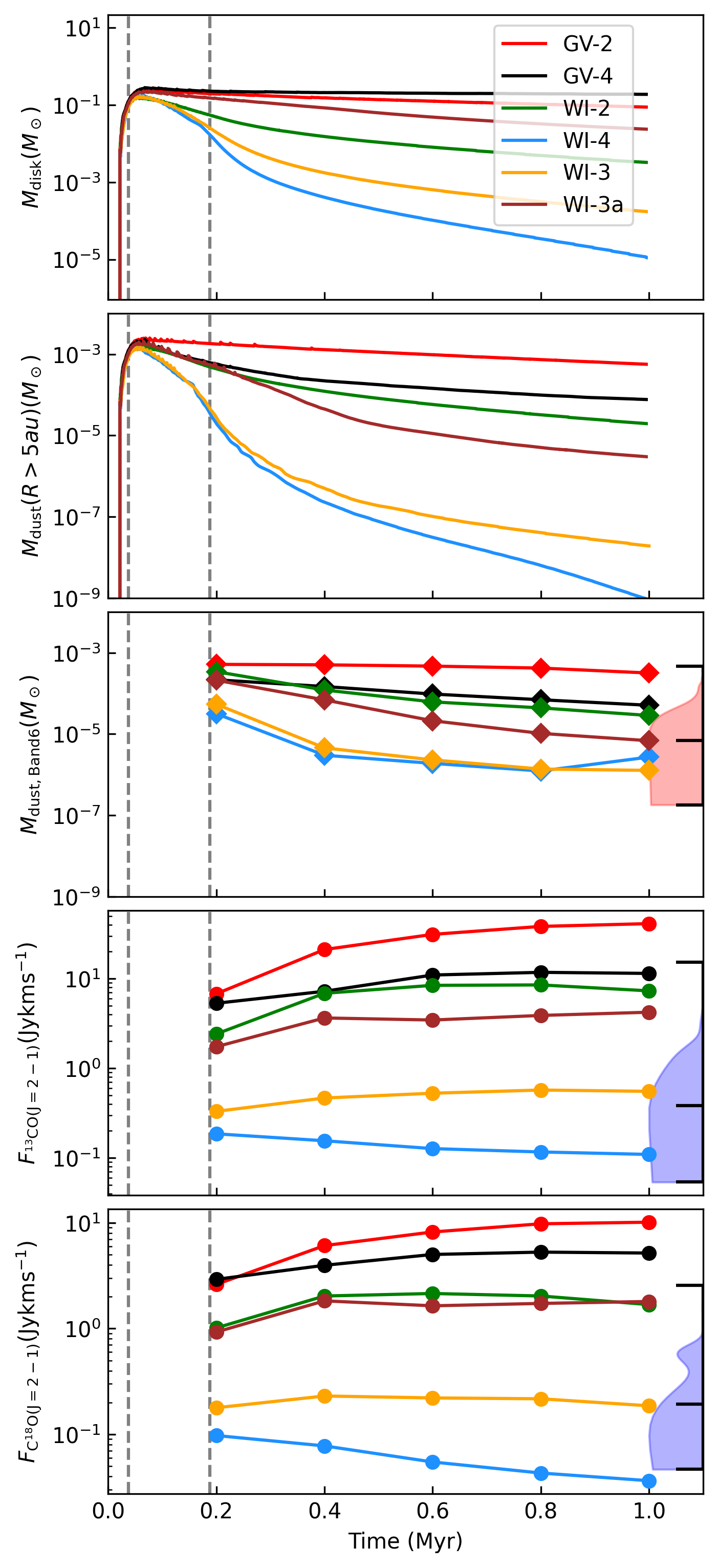}
\caption{Evolution of the disk mass. Top panel: Theoretical mass within the centrifugal region. Second panel: Theoretical dust mass within the centrifugal disk, neglecting the innermost five au region.
Third panel: Dust mass obtained from {\sc ProDiMo} synthetic observations. Last two panels: Synthetic CO line flux as a proxy for the disk gas content.  
The dust masses are calculated from continuum Band 6 fluxes, while the $^{13}$CO and C$^{18}$O line fluxes are calculated at 150 pc.
The vertical, dashed gray lines mark the average Class 0/I and Class I/II boundaries.
The shaded regions on the right depict half-violin plots of dust masses (red) and line fluxes (blue) in Class II disks, as observed in a subset of ALMA surveys. The vertical, dashed gray lines mark the average Class 0/I and Class I/II boundaries.
}
\label{fig:Mdiskevo}
\end{figure}

In Fig. \ref{fig:Rdiskevo}, we observe that for each model, $R_{\rm cf}$ lies between the synthetic gas and dust radii at all times, with the synthetic gas radius consistently larger than $R_{\rm cf}$.
As discussed before, the gravitoviscous disks spread rapidly to large sizes, both in dust and gas, with the disk spread increasing with the viscous $\alpha$-parameter.
The models with magnetic winds produce gas disks that are smaller than corresponding gravitoviscous models.
At low values of $\alpha_{\rm MRI}$, the wind models \simname{WI-3} and \simname{WI-4} result in very small continuum sizes, which lie well outside of observational bounds. 
When the disk winds are active, these low alpha models produce significantly smaller initial Class 0 disks.
The subsequent advective action of winds overpowers the gravitoviscous spread, which results in overall smaller gas disks.

Considering the evolution of dust radius in the third panel of Fig. \ref{fig:Rdiskevo}, the \simname{GV} disks show a monotonic increase.
With the inclusion of winds, the dust disk radius decreases as a disk enters Class II stage and afterward, it may increase depending on the model. 
{Recent survey of Orion molecular clouds indicates a similar decrease in the mean dust disk radius with evolution, which supports the scenario of disk winds \citep{Tobin+20}.
The impact of dust evolution on the disk radius in the presence of winds can be summarized as follows.
The dust size in the outer regions of a PPD is typically drift limited, while in the inner regions the dust may grow to reach the fragmentation barrier \citep{Birnstiel16}. 
As a consequence of the dust partitioning, in the winds in the outer disk evacuate a relatively larger amount of (small) dust per unit mass.
The remaining grown dust is subject to efficient radial drift toward increasing gas pressure in the low $\alpha$ environment, since a lower $\alpha$ increases dust fragmentation barrier, dust maximum size, and the Stokes number. 
PPDs subject to photoevaporative winds cause a reduction of dust reservoir at large radii by a similar mechanism \citep{Sellek+20}. 
The bulk of the disk is thus depleted of small dust and grown dust drifts to the inner disk regions where it gets trapped in the gaseous rings, with a large dust-to-gas ratio.
As the result, the dust disk size in \simname{WI-3} and \simname{WI-4} shrinks to only a few au in Band 6 continuum flux.}
With the adjust wind model \simname{WI-3a}, the effects of wind are attenuated by increasing $C_\beta$ parameter by a factor of 5. 
This modulates the inward transport of gas and consequently the dust drift so that the continuum disk in synthetic observations are congruent with observations even at low $\alpha_{\rm MRI}=10^{-3}$.
The dust dynamics with disk winds will be studied in detail in a subsequent paper.

Figure \ref{fig:Mdiskevo} shows the evolution of disk mass for the models.
The true disk gas mass in the model ($M_{\rm disk}$) is plotted in the first panel, 
which is calculated by summation over the extent of the centrifugally supported disk.
All disks achieve the maximum disk mass at the beginning of the Class I stage and afterward in the Class II stage, the masses decrease monotonically at different rates.
As a consequence, the disk mass is typically a substantial fraction of the stellar mass during the Class I stage and the disk is prone to gravitational instability (see second column of Fig. \ref{fig:global}).
In addition to being large in size, the \simname{GV} models typically produce more massive disks, especially at low viscosity. 
With a low value of $\alpha_{\rm MRI}=10^{-4}$ in model \simname{GV-4}, the transport of angular momentum due to viscous torques is diminished, which over time results in the accumulation of material and a more massive disk than \simname{GV-2}.
Both models \simname{WI-3} and \simname{WI-4} produce disks with masses in the vicinity of $10^{-4} M_\odot$ or 0.1 $M_{\rm J}$. 
Thus, these disks do not have sufficient mass to produce giant planets, which accrete gas over timescale of several million years in the core accretion scenario \citep[e.g.,][]{Pollack96}.
{However, we note that the planet formation possibly begins before Class II stage and at that time, the mass available for this process is much more than typical assumptions of MMSN disk \citep{Tychoniec+20,Schib+21}.}
Models \simname{WI-2} and \simname{WI-3a} produce disks that are of the order of 0.01 $M_\odot$ and are most consistent with our understanding of PPD masses in the T Tauri phase.

{A key trend observed in the first panel of Fig. \ref{fig:Mdiskevo} is that the lower viscosity disks with gravitoviscous evolution are more massive. However, the trend is reversed with the inclusion of disk winds; that is, lower viscosity disks become progressively less massive over time.}
This can be explained with the strong dependence of disk wind mass loss and torque on the gas surface density.
{  The low viscosity creates both mass and angular momentum transport bottleneck, and forces the disk to have a larger $\Sigma_{\rm g}$, which in turn produces strong winds with more torque as well as mass loss.
The net effect is that the winds remain strong, until the surface density is reduced, which ultimately results in a less massive disk.}
This should increase the mass of the central star in wind models { relatively rapidly}, which is seen in Table \ref{table:final}.

{ 
The second panel of Fig. \ref{fig:Mdiskevo} shows the theoretical dust mass of the disk, calculated as the sum of both small and grown dust components.
This is done with an exclusion of the innermost five au of the disk, in order to avoid the region where the disk shows formation of dusty rings \citep{Kadam+19,Kadam+22}. 
These rings form pressure maxima and tend to foster dust growth up to meter-sized boulders, which are not observable and contain significant amount of dust mass \citep{Vorobyov+24}.
Additionally, the innermost region is typically optically thick, and does not reveal its true mass in observed disks.  
Thus, removing the innermost region from our theoretical dust mass estimates seems to be the best compromise, for comparison with the synthetic observations.
For models \simname{WI-3} and \simname{WI-4}, the dust radius collapses after about 0.5 Myr (last panel of Fig. \ref{fig:Rdiskevo}), implying that the calculated dust masses outside of five au are correspondingly small and not representative of the dust content of the disk.
However, for the rest of the models, the theoretical dust masses serve as a good proxy for the bulk of the observable dust mass in the disk.
}

The third panel of Fig. \ref{fig:Mdiskevo} shows the synthetically derived dust masses of the simulated disks in comparison with observations.
The red violin plot in this panel shows the observed distribution of dust masses as calculated from ALMA Band 6 continuum data, where all four nearby star forming regions are considered \citep{Miotello+23,Manara+23}.
The abrupt cutoff at the lower end of the observed kernel density distribution is caused by the detection limit of the surveys. 
Similar to the ALMA surveys, we assume that the dust emitting at submillimeter wavelengths is optically thin and it emits isothermally at a characteristic temperature of $T_{\rm d}=20$ K \citep{Hildebrand83,Andrews-Williams05,Manara+23}.
This allows us to estimate the disk mass from the Band 6 (1.3 mm) flux ($F_{\nu}$) as
\begin{equation}
    M_{\rm dust, Band 6} = \frac{F_{\nu} d^2}{ \kappa_{\nu} B_{\nu}(T_{\rm d})},
    \label{eq:Mdustobs}
\end{equation}
{where, $B_{\nu}(T_{\rm d})$ is the Planck function at a dust temperature of $T_{\rm d}=20$ K, and $\kappa_\nu$ is an average dust grain opacity, calculated as $\kappa_\nu = 3.5 (870 \mu {\rm m} /\lambda)$ ${\rm cm^{2}g^{-1}}$ \citep{Beckwith+90, Miotello+23}.
Considering the masses at the Class 0/I boundary, the synthetic dust masses at this early times are consistently larger than those indicated by recent survey ($\approx 10^{-5} - 10^{-4} M_\odot$ for Orion molecular clouds \citep{Tobin+20}), except in the case of \simname{WI-3} and \simname{WI-4}.
When compared against the observed distribution of dust masses of Class II disks, \simname{GV-2} results in a massive disk at the upper end of the observed range.}
The significantly larger gravitoviscously spreading disk in \simname{GV-2} model contains the largest amount of dust mass.
We find that the dust masses inferred from {\sc ProDiMo} fluxes for all other disks are consistent with the observed range of the surveys, with low viscosity disks with winds producing a better fit.
{ Comparing with the theoretical dust masses in the second panel, the synthetic disk dust masses show an excellent agreement for all models throughout the Class II evolution. As expected, \simname{WI-3} and \simname{WI-4} diverge after about 0.5 Myr, for aforementioned reasons.}

As mentioned earlier, gas mass of a PPD may be inferred by multiplying its dust mass by a factor of 100.
If we calculate the gas masses this way, the estimates will follow the exact same trends as the third panel of Fig. \ref{fig:Mdiskevo}, shifted up by the same factor.
As estimating disk gas mass comes with significant uncertainties, for the most transparent and direct comparison against the observations, the last two panels of Fig. \ref{fig:Mdiskevo} show the $^{13}$CO and C$^{18}$O line fluxes at 150 au. 
The observed ranges are obtained from ALMA survey data for Lupus and Chamaeleon star forming regions \citep{Ansdell+16,Long+17}. 
In general, the line fluxes from the models are larger than the observed distributions and there may be several contributing factors.
Firstly, we note that in {\sc ProDiMo}, CO isotope-selective processes are not considered and {the abundance of CO isotopologues is inferred with fixed ratios of $[^{13}{\rm C}]/[^{12}{\rm C}]=0.0140$ and $[^{18}{\rm O}]/[^{16}{\rm O}]=0.0020$.}
The less abundant CO isotopologues are thought to become optically thick much deeper in a PPD, thereby tracing the entire thickness of bulk of the disk.
However, the increase in line flux with time for the higher mass models suggests that both of these traces are optically thick and therefore follow the increase in disk size with gravitoviscous spread.
The simulated disks evolve only up to one Myr, while the considered star forming regions may be up three Myr old. 
As a PPD looses its mass with age, this would decrease the observed CO fluxes with respect to synthetic observations.
The {\sc ProDiMo} fluxes are calculated for a face-on disk with zero inclination, which may contribute to a higher CO flux as compared with the observed sample of disks.
Finally, we note that CO observations imply a bulk gas-to-dust ratio between 1-10 for most disks, which is much lower than canonical MMSN or ISM value of 100 \citep{Miotello+17}.
This discrepancy is attributed to carbon depletion in PPDs, with processes such as CO photodissociation in the upper layers and freeze-out near the midplane. 
{If the CO emissions are optically thick and dust emission optically thin, this will also result in an underestimation of the measured bulk gas-to-dust ratio as compared to the actual value.}  
{ The CO fluxes demonstrate that the adjusted disk wind model does not fit all observables simultaneously. }
Nonetheless, with the inclusion of disk winds, the simulated line fluxes decrease and move in the correct direction with respect to an agreement with the observed distributions.

\begin{figure}
\centering
  \includegraphics[width=0.5\textwidth]{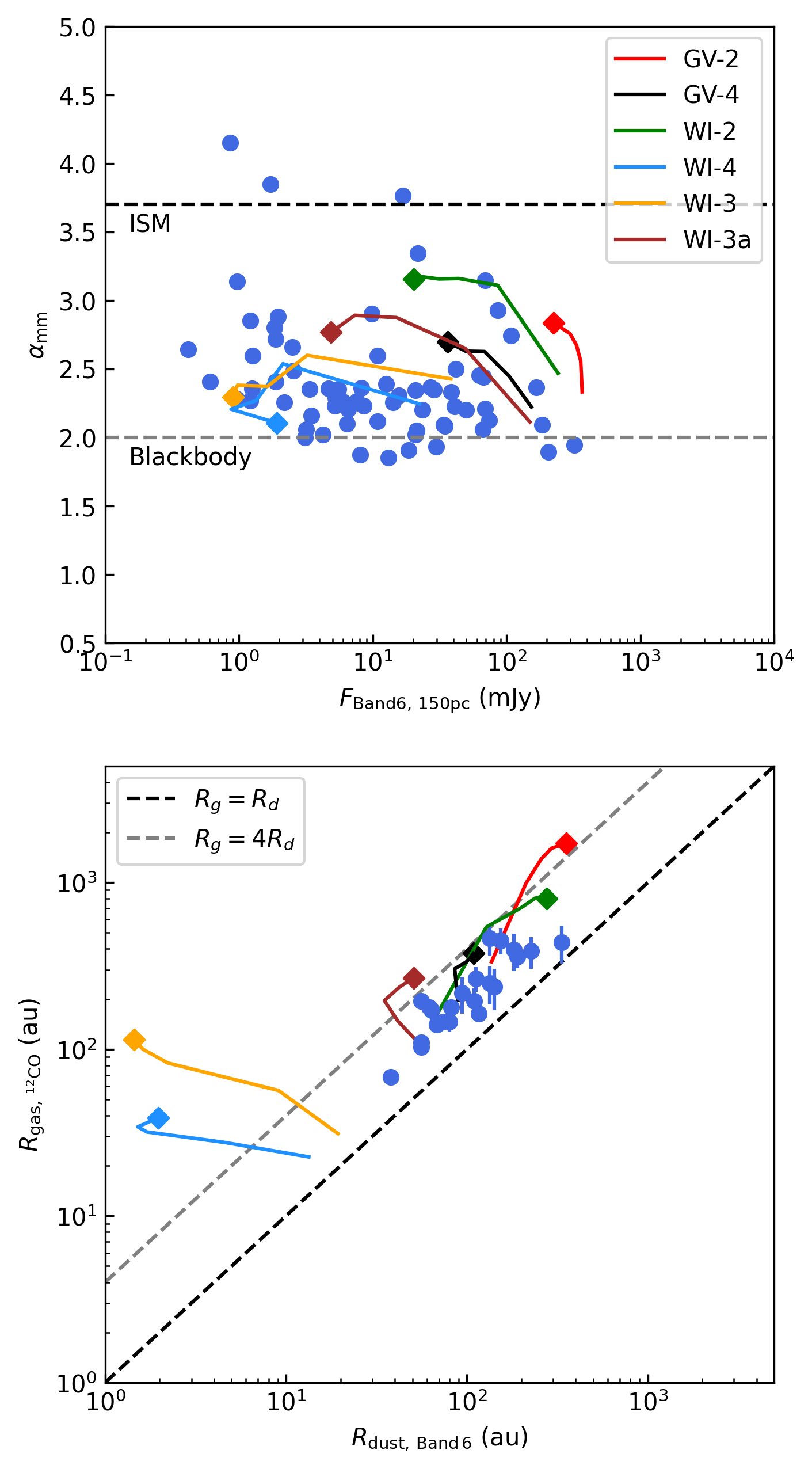}
\caption{Comparison of the synthetic observations obtained from the disk models with the ALMA survey data with respect to the spectral indices (first panel) and dust-gas radii (second panel). 
The colored lines show evolutionary sequence of the models, starting at 0.2 Myr and ending at one Myr, with the latter point marked with a diamond.
The blue scatter plots show observed quantities in Lupus cluster \citep{Ansdell+18}. 
}
\label{fig:observed}
\end{figure}

We now test the disk evolution predicted by our models against observed spectral indices and the ratio of dust and gas disk sizes.
Figure \ref{fig:observed} compares the disk models with ALMA survey data of Class II PPDs in Lupus cluster \citep{Ansdell+18}.
The top panel in Fig. \ref{fig:observed} shows the total continuum flux versus far infrared spectral index between ALMA Band 6 (1.3 mm) and 7 (0.9 mm).
The latter is the SED slope defined as
\begin{equation}
    \alpha_{\rm mm} = \frac{{\rm log}(F_{\rm 0.9 mm}/F_{\rm 1.3 mm})}{{\rm log}(\nu_{\rm 0.9 mm}/\nu_{\rm 1.3 mm})}, 
    \label{eq:spectralindex}
\end{equation}
where $F_\nu$ is the continuum flux at the corresponding ALMA band.
{The $\alpha_{\rm mm} = 2$ line is characteristic of large particles represented by blackbody emission, since $F_\nu \propto \nu^2$. 
The ISM value of 3.7 corresponding to small particles is also shown in the plot.}
Assuming optically thin emission, the flux in submillimeter continuum is a direct proxy of the total dust content in the disk.
{The spectral index probes the characteristic grain size and is hence often called the ``growth index."}
Although there is a higher degree of uncertainty, as additional factors such as optical depth and dust scattering complicate the picture \citep{Draine06}. 
{\sc ProDiMo} calculates the fluxes at ALMA Bands at a distance of 150 pc and the spectral index is calculated from Eq. \eqref{eq:spectralindex}.
We note that the observations have large error bars in spectral index, which are not shown in the figure.
A disk starts at 0.2 Myr, when it is fully formed and typically evolves diagonally from lower right to upper left this phase space (diamond in the figure represents the end state at one Myr). 
Similar trends in the disk evolution have been previously shown in several studies \citep{Birnstiel+10,Pinilla+12,Akimkin+20}. 
{The decrease in Band 6 flux indicates a decrease in dust mass or, in the case of optically thick emissions, an area effect due to inward dust drift. }
The evolution of spectral index is more complicated; the initial spectral index near 2 is a result of initial large optical depths and its value increases when the grain size approaches $\approx \lambda /2 \pi$ of the corresponding wavelength.
The disks with large viscous $\alpha$ tend to show both larger flux and spectral index.
The values of both spectral index and continuum flux obtained from the models are consistent with those in the ALMA surveys throughout the first Myr of evolution. 
The $\alpha_{\rm mm}$ for \simname{WI-3} and \simname{WI-4} models first increases and then decreases with time.
{This decrease at later evolutionary stages is very likely a signature of dust clearing process, as all big grains drift inward to the pressure maxima (see Sect. \ref{subsec:synthobs}).
The $\alpha_{\rm mm}$ is essentially disk-averaged spectral index and corresponds to the most luminous parts of the disk.
As most of the dust mass in these low $\alpha$ models with winds eventually accumulates in the inner rings, the associated increase in the average optical depth may be responsible for the re-decrease in $\alpha_{\rm mm}$ at later times. }

The second panel of Fig. \ref{fig:observed} compares the co-evolution of the gas and dust radii. 
These radii obtained from {\sc ProDiMo} continuum emissions are same as those presented in Fig. \ref{fig:Rdiskevo}.
The observations suggest that a PPD has its gas radius larger than the dust radius by up to a factor of 4, as indicated by the diagonal dashed lines in the figure.
Similarly, the disk radii inferred from simulations consistently show $R_{\rm gas} > R_{\rm dust}$.
As elaborated in Sect. \ref{subsec:synthobs}, a smaller dust radius can be attributed to dust growth and drift and it is also related to a drop in the dust opacity coefficient.
The models with high $\alpha_{\rm MRI}$, \simname{GV-2} and \simname{WI-2} show evolution toward top right, as a consequence of gravitoviscous spreading. 
The low $\alpha$ models with winds, \simname{WI-3} and \simname{WI-4}, deviate significantly from the observed trends.
As discussed in the context of Fig. \ref{fig:Rdiskevo}, the overpowering magnitude of the winds in this case results in small disks, and increased dust drift results in an increasingly compact disk in Band 6 continuum. 
We note that the only difference between models \simname{WI-2}, \simname{WI-3} and \simname{WI-4} is viscous $\alpha$.
Thus, when the disk winds are involved, the exact magnitude of turbulence plays a significant role in overall disk evolution.
The adjusted wind model \simname{WI-3a} is much more consistent with the observational limits, as compared to \simname{WI-3}. 
However, it still leaves some room for improvement, as the ratio of gas to dust radius is somewhat larger than $R_{\rm g} = 4 R_{\rm d}$ line.
We note that the aim of model \simname{WI-3a} is not to fit this wind model perfectly to the observations. 
Rather we adjust this model to show trends of varying the parameters associated with disk winds ($C_\beta$ and $C_H$) and demonstrate that a disk with reasonable characteristics can be obtained at low alpha without fine tuning.

In Table \ref{table:final}, we list some of the global disk quantities that characterize the disks at one Myr.
The numbers reinforce several conclusions made earlier and here we discuss some of the key points in terms of the mass budget of the PPDs.
Model \simname{GV-4} produces a massive disk capable of forming envelopes of hundreds of gas giants, while \simname{WI-4} and \simname{WI-3} do not have enough mass to form a single gas giant. 
Most of the dust mass is in the grown dust, which is expected from dust dynamics occurring in a PPD.
The low viscosity increases the fragmentation barrier for dust grains and the disk winds increase the inward dust drift, resulting in much more grown dust mass as compared to the small dust in these models.
{The global bulk dust to gas ratio ($\zeta_{\rm d2g}=M_{\rm disk, dust}/M_{\rm disk, gas}$) is approximately between $0.1-5 \%$ and in most cases, the dust is depleted below the canonical ISM value of $1\%$. }
The ``efficiency" of star formation can be inferred from the stellar mass, as all models start from the same mass reservoir of 0.83 $M_\odot$. 
With gravitoviscous evolution, the lower $\alpha$ models produce a star with lower mass, as the diminished angular momentum transport maintains a relatively massive disk.
With the inclusion of winds, however, we see that the efficiency of star formation is almost independent of disk $\alpha$ and varies with the strength of the winds. 
The total mass lost to the winds is inversely proportional to viscous $\alpha$, as these disks produce stronger winds via its dependence on the gas surface density. 
{However, a larger amount of net mass lost to the winds does not necessarily mean a less massive disk.
For example, in the case of \simname{WI-3a}, the wind stress is attenuated by increasing $C_\beta$, while the wind mass loss rate is increased by increasing $C_H$.
Comparing against \simname{WI-3}, the disk mass is larger, while the mass lost to the winds comes at the cost of the final mass of the star.}
Finally, the $t_{\rm disk}=M_{\rm gas}$/$\dot{M}_\star$, where $\dot{M}_\star$ is stellar accretion rate, is calculated at one Myr. 
This rough estimate of the disk lifetime implies that the disks with magnetic winds evolve faster with a noticeably shorter lifespan.

\begin{table*}
\caption{Comparison of global disk properties at one Myr.}
\label{table:final}
\begin{tabular}{p{1.0cm}p{1.2cm}p{1.2cm}p{1.2cm}p{0.8cm}p{0.8cm}p{1.4cm}p{1.2cm}p{1.2cm}p{1.3cm}p{0.5cm}p{0.5cm}p{0.6cm}}
\hline
Model  &   $M_{\rm gas} $ &  $M_{\rm dust,gr} $   &  $M_{\rm dust, sm} $  &  $\zeta_{\rm d2g}$   &  $M_{\rm star} $ & $M_{\rm disk}/M_{\rm star}$ &  $M_{\rm w,lost} $ &   $R_{\rm gas, ^{12}CO }$    &  $R_{\rm dust, 1.3 mm}$  &  $R_{\rm cf}$     &     $\lambda$  & $t_{\rm disk}$  \\ 
  &$(10^{-2} M_{\odot} )$&  $( 10^{-5} M_{\odot} )$   &  $( 10^{-5} M_{\odot} )$  &  (\%)   &  $(M_\odot)$ &  (\%)  &  $( 10^{-2} M_{\odot} )$ &    (au)   &   (au) &   (au)     &       & (Myr)  \\
\hline
\simname{GV-2}   &   9.987  &  55.96 &  13.42  &  0.69    &   0.64   & 11.7 &  9.51 & 1720 & 355 & 481 & 2.6 &  2.0 \\
\simname{GV-4}   &  21.005  &  19.82  &  0.96  &   0.10   &   0.56   & 37.4 &  10.86 & 376  & 109 & 256 & 5.3 & 3.2 \\
\hline
\simname{WI-2}   &   0.458   &  20.00  &  1.14 &   5.3   &   0.69   & 0.58 & 6.22 & 824 & 276 & 156 & 1.8 &  0.83  \\
\simname{WI-4}   &  0.005  &  5.4  &  0.03  &   1.1   &   0.70  & $6.9 \times 10^{-5}$ & 6.16 & 39 & 2.0 & 5.7 &  2.0 &  0.57  \\
\simname{WI-3}   &   0.026  &  6.61 &  0.03  &   0.25   &   0.69    &  $3.7 \times 10^{-4}$  & 6.19 & 114 & 1.4 & 26 & 1.3 & 0.77 \\
\simname{WI-3a}   &  2.902  &  8.53 &  0.11  &  0.29    &   0.64    &  4.5 & 9.25 & 268 & 51 & 226 & 1.2  & 0.98  \\
\hline
\end{tabular}
\end{table*}

\subsection{Theoretical expectations from disk winds}
\label{subsec:theoryexp}

\begin{figure}
\centering
  \includegraphics[width=0.5\textwidth]{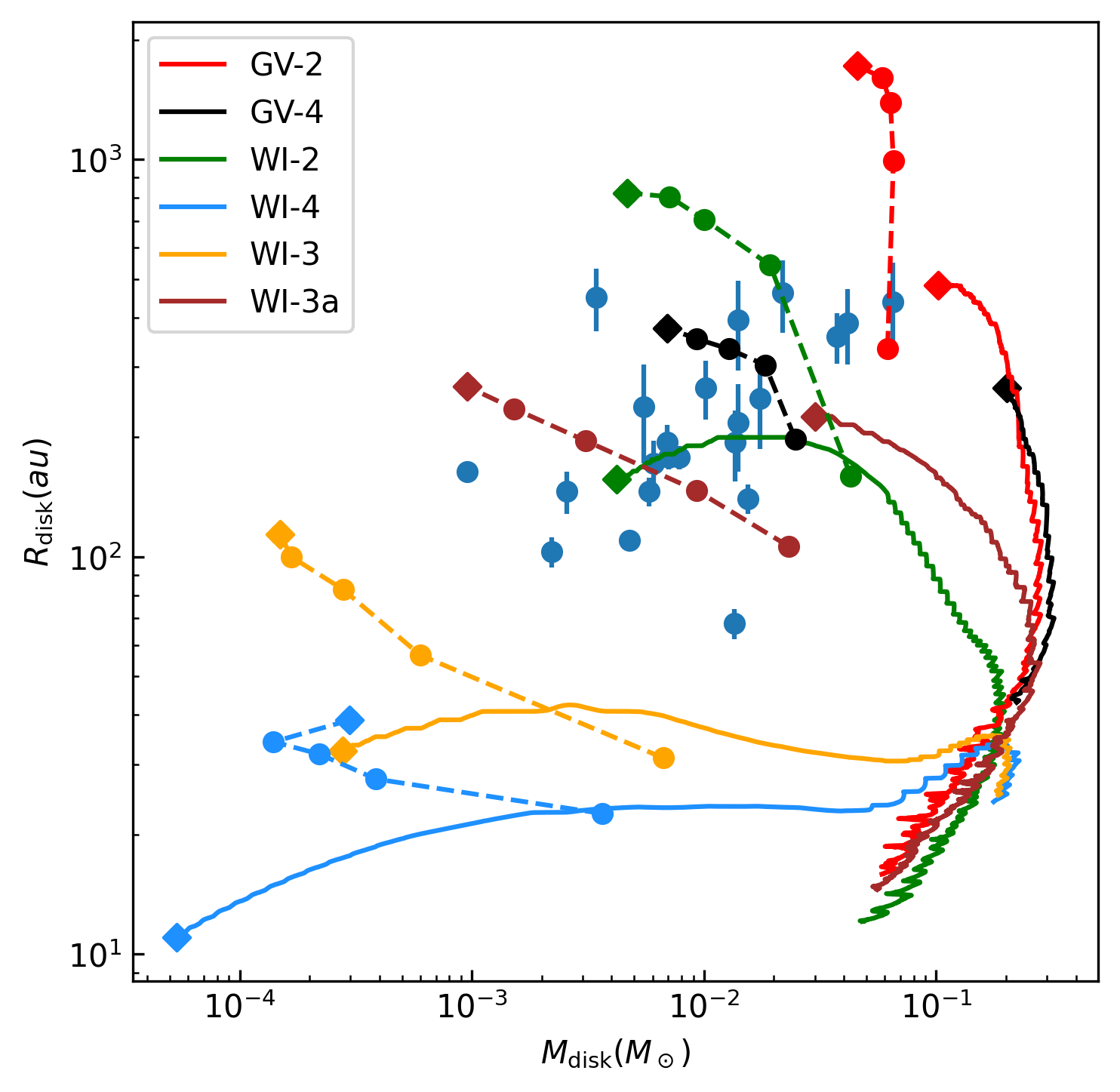}
\caption{Evolution of the disk models in the mass-radius plane showing how the wind models may diverge from the gravitoviscous models after the disk achieves its maximum mass. The solid lines depict theoretical disk size, $R_{\rm cf}$, versus gas mass within the centrifugal disk. { The dashed lines show $R_{\rm gas, ^{12}CO}$ obtained from synthetic emissions versus disk mass, the latter calculated as 100 times the dust mass obtained from the synthetic continuum emission.} The diamonds mark the end points at one Myr and the blue scatter plot shows observed values from an ALMA survey of the Lupus cluster \citep{Ansdell+16,Miotello+23}.
}
\label{fig:RM}
\end{figure}

In a recent review on protoplanetary disk demographics, \cite{PP7-15} illustrate some potential population level differences that can distinguish disks that evolve via predominantly viscous processes versus those driven primarily by disk winds.
{This difference arises because the action of magnetic winds tends to shrink the disk size as opposed to its viscous spreading, and it is best distinguished via the divergence of the evolutionary tracks in the mass-radius plane.
Figure \ref{fig:RM} depicts such a plot for the simulated models. Here the solid lines depict the theoretical centrifugal radius plotted against the disk mass, the latter corresponding to the gas mass enclosed within the centrifugal disk.}
$R_{\rm cf}$ is smoothed in time with a moving boxcar average with a box size of 0.05 Myr, hence these quantities correspond to the first rows of Figs. \ref{fig:Rdiskevo} and \ref{fig:Mdiskevo}.
The dashed lines show the disk sizes in $^{12}$CO, obtained from synthetic emissions and plotted against the { synthetic disk mass at that time. The latter is calculated as 100 times the dust mass derived from the continuum emissions, since many studies still use this approximation.
The diamonds mark endpoints of the simulations at one Myr.}
The scatter plot corresponds to the observed disks in Lupus cluster as seen in the ALMA survey data, where the disk masses are obtained from the Band 6 dust content \citep{Ansdell+16,Miotello+23}.
The region on the extreme right represents the early and embedded GI-prone state when the disk is massive, while the trajectory will eventually move left toward lower mass and eventual dispersal, presumably via photoevaporation.
The larger observed radius as compared to $R_{\rm cf}$ is seen in almost all models, which may be an artifact of a 90\% flux threshold or additionally, of the criterion of surface density threshold in calculating $R_{\rm cf}$.
{ On the other hand, the observed disk masses are systematically smaller, which are a result of significant dust depletion as well as growth during the disk evolution.}

{The gravitoviscous models without disk winds spread monotonically, and end up with either a large disk radius (\simname{GV-2}) or a large disk mass (\simname{GV-4}).
{ The low viscosity GV-4 model is able to hide most of its gas mass; however, such massive disks, which are a significant fraction of the stellar mass, typically exhibit GI spirals.}
Thus, GV models may be congruent with the observations of the largest or the most massive of the Class II disks.}
When both gravitoviscous and wind-driven accretion are present, the disk spreading is reduced and the radius evolution is essentially a result of the tug-of-war between these two processes.
Hence, with the inclusion of disk winds, PPDs tend to evolve toward the diagonal in Fig. \ref{fig:RM} so that they end up smaller and less massive.
{The disk wind models with low alpha, \simname{WI-3} and \simname{WI-4}, produce compact disks that lie outside of observational expectations, although since $R_{\rm gas, ^{12}CO} > R_{\rm cf}$, model \simname{WI-3} moves toward the observed values.}
Model \simname{WI-2} with a larger value of $\alpha$ as well as \simname{WI-3a} are more or less congruent with the observed ranges. 
$R_{\rm cf}$ in \simname{WI-2} contracts near the end of its evolution, implying the size does not necessarily increase monotonically when disk winds are considered, although the synthetic disk size in $R_{\rm gas, ^{12}CO}$ increases with time.
These results are qualitatively consistent with the expectations from previous models of disk winds, and imply that magnetic disk winds may be necessary to explain observed distributions \citep[e.g.,][]{Bai+16,Tabone+22}.
Figure \ref{fig:RM} demonstrates that in reality, both the measurements of disk sizes and masses contain several assumptions and significant uncertainties to draw decisive conclusions about the disk driving mechanism.

At this point, we discuss the dimensionless quantity of magnetic lever arm, $\lambda$, which is typically assumed as a parameter for wind-driven accretion models that are analogous to simple viscous $\alpha$-disks \citep{Kimmig+20, Alessi-Pudritz22, Tabone+22}. 
This parameter characterizes the removal of angular momentum and it is essentially the ratio of specific angular momentum evacuated by the wind to that in the underlying Keplerian region that launches the wind. 
In the classical picture of magnetocentrifugal winds, the magnetic field lines rotate rigidly inside of the Alfv\'{e}n surface so that the gas leaving as a wind increases its specific angular momentum \citep{Blandford-Payne82}.
If a field line anchored at a radial distance $r_0$ crosses the Alf\'{v}en surface at $r_A$, the magnetic lever arm can be written as 
\begin{equation}
\lambda= \bigg(\frac{r_A}{r_0} \bigg)^2 = \frac{1}{2} \frac{\dot{\Sigma}_{\rm w,tr}}{\dot{\Sigma}_{\rm w,g}},
\label{eq:lever}
\end{equation}
where $\Sigma_{\rm w,tr}$ is the radial mass transport rate resulting from the magnetic winds. 
The lever arm decides how efficient the disk winds are in angular momentum removal, since with a larger $\lambda$, a small amount of mass lost carries a relatively large amount of angular momentum.
In general, this parameter has to be greater than unity in order to extract angular momentum and brake the disk.
Recent MHD simulations of weakly magnetized disks suggest $\lambda$ to be between 1.1 and 5, with a typical value of $1.3$ \citep{WBG19,Gressel+20,Lesur21}. 
Less efficient winds at $\lambda \approx 1$ may primarily be driven by vertical gradients of magnetic pressure.
Conical molecular outflows have been observed in a handful of sources, which give estimates for $\lambda$ that are consistent with the simulations \citep{deValon+22,Louvet+18,Tabone+17}.  

\begin{figure}
\centering
  \includegraphics[width=0.5\textwidth]{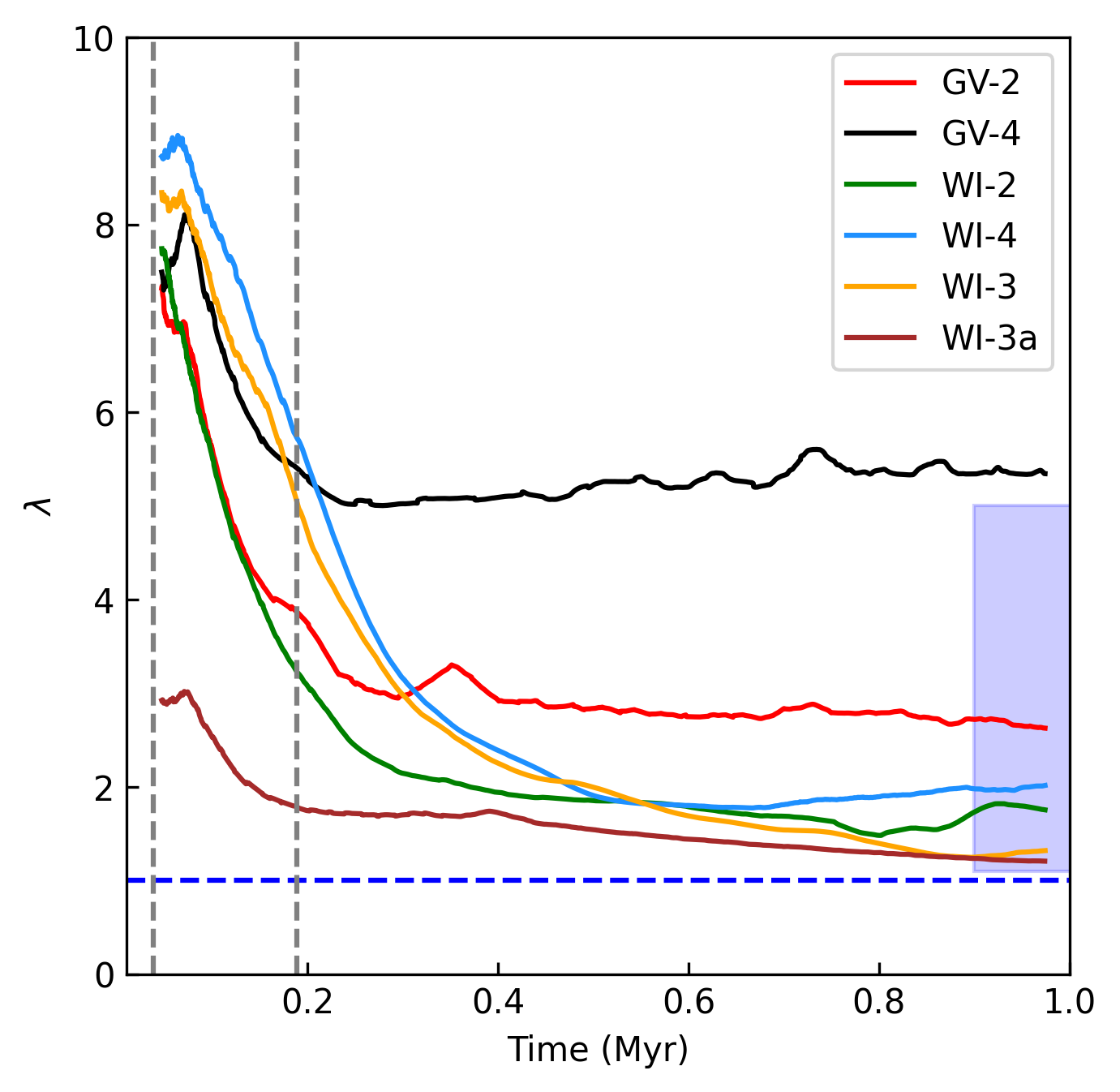}
\caption{Evolution of the inferred value of the effective lever arm for the disk models. The vertical, dashed gray lines mark the average Class 0/I and Class I/II boundaries and blue, dashed line indicates lever arm length of unity. The shaded region on the right shows the expected range of values.}
\label{fig:lever}
\end{figure}

In our disk wind model, both the mass loss and angular momentum loss rates depend on local disk quantities and stellar properties, while a lever arm parameter is not used explicitly. 
However, it is possible to infer the value of $\lambda$ by calculating it a posteriori from Eq. \eqref{eq:lever}.
Assuming steady state accretion and vertically symmetrical wind geometry,
\begin{equation}
    \dot{\Sigma}_{\rm w,tr} = \frac{8 \pi}{\Omega} r T_{z\phi}^{\rm Max}.
    \label{eq:mdotWtr}
\end{equation}
In the simulations, the local values of $\lambda$ fluctuate severely. 
However, a global $\lambda$ can be calculated, which is averaged over the entire centrifugal disk at a given time.
Figure \ref{fig:lever} shows the evolution of this inferred $\lambda$, which is smoothed with moving boxcar average with a box length of 0.05 Myr.
For all models, $\lambda$ decreases initially, especially during the Class I phase, while its value remains remarkably constant at later times. 
During Class 0 and I stages, the disk is massive and also relatively compact, which results in large average gas surface densities. 
This implies greater wind torques at early times and a correspondingly large values of $\lambda$.
For \simname{GV-2} and \simname{GV-4}, where disk winds are not included self-consistently, $\lambda$ is relatively large at later times as well. This may be a result of a larger average $\Sigma_{\rm g}$, as disk winds are not included self-consistently in the \simname{GV} models.
For the disk wind models, the lever arms converge to $\lambda \approx 1-2$, which is in excellent agreement with both expectations from MHD simulations and observational constraints. 
The global values of $\lambda$ are relatively steady during Class II stage in all cases.
This suggests that simpler disk wind models that start with an MMSN disk and adopt a constant lever arm may also be suitable for long-term evolution.

\section{Discussions and conclusions}
\label{sec:conclusions}

Magnetic disk winds play a crucial role in the evolution of protoplanetary disks, as suggested by both observational evidence and theoretical insights over the past two decades.
In this study, we have introduced a simple semi-analytical formulation of magnetic disk winds in the thin-disk limit. 
{The disk wind model is based on the insights gained from local shearing box simulations. The results of such local simulations can be stitched together to obtain analytical radial dependence of the wind properties throughout the disk \citep{Bai13}.} 
These fitting formulae assume a power-law dependence of wind mass loss rate and surface stress as a function of local disk properties.
We improved upon these results and took into account several corrections, for example, due to changing stellar mass and FUV luminosity.
We incorporated this model of magnetic disk winds self-consistently in the FEOSAD code, which simulates the formation of protoplanetary disks and their evolution over the long term. 

The wind mass loss rate and torque result in a sink term in mass continuity and momentum equations, respectively, while acting over the extent of the centrifugal disk.
The corrections to other relevant equations, such as energy conservation and those for small dust, were also considered. 
The two free parameters in the disk wind model are $C_\beta$, which arises from the ideal MHD approximation used for the disk evolution, and $C_{\rm H}$, which originates in the relatively unconstrained mass loss rate as predicted by the shearing box simulations.
{ 
The simulation results depend sensitively on the factor $C_\beta$, which scales the disk plasma $\beta_0$ in our ideal MHD models. However, we note that this parameter is not arbitrary, as it is chosen so that the quantity $\beta_0 \times C_\beta$ over majority of the disk is of the order of $ 10^{4}$. 
This expected value is consistent with $\beta_0$ in protoplanetary disks from both observations and simulations, although these estimates vary by orders of magnitude. 
If our model is correct, only a narrow range of $C_\beta$ may be suitable for disk evolution, implying that there could be an underlying mechanism -- perhaps some self-regulation process during magnetic transport -- that causes different protoplanetary disks to have similar magnetic $\beta_0$ values.
Moreover, $C_\beta$ may not be constant in space and time, and this uncertainty makes it challenging to compare models against observations. 
For the model to work without $C_\beta$, solving the non-ideal MHD equations is necessary.
{  An alternative approach could be capping $B_z$ to a reasonable upper limit corresponding to observational constraints, for example, from paleomagnetism of meteorites in the early Solar System \citep{Fu+14,Borlina+21}. 
This approximation handles the two limiting cases of no and strong diffusivity correctly, as $B_z$ will be approximately constant in the limit of strong magnetic diffusion. 
Despite the aforementioned drawbacks, our model gives us a reasonable insight into the long-term evolution of wind-driven disks.}

}

Previously, FEOSAD considered gravitoviscous evolution, wherein a PPD forms from the collapsing parent cloud and subsequently evolves via combined action of turbulent viscosity and gravitational torques.
{In this framework, we conducted numerical experiments that incorporated the effects of disk winds.}
In our limited parameter space study, three values of viscous $\alpha$ were considered ($0.01$, $10^{-3}$, and $10^{-4}$). 
Synthetic observations were obtained by post-processing the simulation results with the radiation thermo-chemical code {\sc ProDiMo}.
For comparison with observations, we aggregated available ALMA survey data for nearby star forming regions.
We validated our magnetic disk wind model with respect to observational constraints, while the disk properties and characteristics of the magnetic winds will be investigated in detail in a subsequent paper.
Without discussing the individual models in detail, our salient findings can be summarized as follows:
\vspace{-0.2cm}
\begin{enumerate}[1]
    \item  Protoplanetary disks evolving under gravitoviscous torques only (without disk winds) typically result in Class II disks that are {too large in mass, and often in size, in comparison with observational expectations. }
    \item With the inclusion of magnetic disk winds, the resulting disks are smaller and less massive. 
    This is expected, as advective evolution caused by vertical angular momentum transport from disk winds tends to suppress the outward gravitoviscous spread and promote inward mass transport.
    \item {Some of the models with winds produce Class II disks that are unreasonably small in terms of dust disk size as well as the true disk mass.
    Obtaining a reasonable fit to the observations is possible with a few adjustments in the two disk wind model parameters $C_\beta$ and $C_{\rm H}$.}
    \item The synthetic observations of global disk properties - gas radius in $^{12}$CO, dust radius in Band 6 continuum, dust content as calculated by Band 6 flux, and $^{13}$CO and C$^{18}$O line fluxes - move in the direction of better agreement with the ALMA survey data when disk winds are included.  
    \item Similar to previous gravitoviscous simulations, models including winds also show the formation of a dead zone, pressure maxima, and dusty rings with an enhanced dust-to-gas ratio in the inner disk region. 
    These can assist rapid planetesimal formation and point toward planet formation starting early in the Class I stage.
    \item{The divergence of evolutionary tracks with respect to disk winds may be seen on the disk mass versus radius plane, 
    where careful choice of disk wind parameters leads to models that are more consistent with the observed ranges.}
    \item 
    The ratio of synthetic gas to dust radii stays within the observed ranges of one to four for the gravitoviscous models.
    At a low viscous $\alpha$, the models that include disk winds may produce anomalously small dust disks due to the loss of small dust and efficient migration of grown dust.    
    By changing model parameters, the magnetic winds can be attenuated so that the resulting disk fits the observational constraints.
     \item The synthetic spectral indices for all models are broadly consistent with the survey data throughout their evolution. As a disk evolves, an increasing $\alpha_{\rm mm}$ indicates grain growth, and a decrease in Band 6 flux indicates a decrease in the dust content.
    \item In agreement with simulations and observations, the inferred effective magnetic lever arm for an entire disk is between approximately one and three for the models that include the action of disk winds, and it stays remarkably steady during the Class II stage. 
\end{enumerate}
\vspace{-0.2cm}

In conclusion, {our magnetic disk wind model is able to reproduce the observational properties while simultaneously satisfying theoretical expectations} over the long term (over one million years), wherein the disk forms and evolves through the Class 0, I, and II stages.
Even now, most models of planet formation start with a fully formed MMSN protoplanetary disk and consider only its viscous evolution.
Recent observations of PPDs put severe constraints on planet formation, as Class II disks do not contain enough dust mass to support the observed population of exoplanets \citep{Greaves-Rice10,Tychoniec+20}. 
{This suggests that planetesimals and planetary embryos start forming earlier in the embedded stages, when the gravitational torques are significant, while disk winds control the fundamental disk properties, such as its size and the mass budget, affecting their subsequent evolution.}
Our model marks a step forward in the direction of representing a more complete picture of such disks, wherein the concurrent effects from viscous, gravitational, and magnetic disk wind processes guide the disk evolution and ultimately decide the fate of the planetary system that emerges.

\begin{acknowledgements}
      We are thankful to the anonymous referee for constructive comments and suggestions. We thank Aleksandr Skliarevskii and Indrani Das for useful discussions. 
      This work utilized computing resources provided by the Digital Research Alliance of Canada. 
      E. I. V. acknowledges support by the FWF project I4311-N27.
\end{acknowledgements}

%
%

\bibliographystyle{aa}
\bibliography{references}

\appendix
\section{Escape velocity of disk winds}
\label{app:windvel}
The escape velocity of the wind, $v_w$, constrains the relative strength of the wind mass loss rate and wind stress.
It can be estimated to the first order by assuming that the work done by the wind torque equals the kinetic energy gained by the mass launched, resulting in 
\begin{equation}\label{eq:vw}
   v_w = \bigg(\frac{2 \tau_w d\theta}{dM_w}\bigg)^{0.5},  
\end{equation}
where $dM_w$ is the total mass lost to the wind (small dust and gas) during a time step. 
The angular distance traveled by the gas parcel during the time step can be calculated as 
\begin{equation}
   d\theta = {\rm arctan} \bigg( \frac{v_{\phi, g} dt}{R} \bigg),
\end{equation}
where $v_{\phi, g}$ is the velocity of the gas parcel in the azimuthal direction.
The wind torque is given by
\begin{equation}
\label{eq:tau_w}
   \tau_w = \frac{1}{4 \pi}  T_{z \phi}^{\rm Max} R dA, 
\end{equation}
where $dA$ is the surface area of the grid cell.
Thus, a moment arm length equal to the local radius is assumed.
With the small angle approximation, 
Eq. \eqref{eq:vw} simplifies to
\begin{equation}
   v_w = \bigg(\frac{1}{2 \pi} \frac{T_{z\phi}^{\rm Max}}{\dot{\Sigma}_{w,g}} v_\phi \bigg)^{0.5}.  
\end{equation}
{Since the ratio ${T_{z\phi}^{\rm Max}}/{\dot{\Sigma}_{w,g}}\approx c_s$, the wind velocity should also typically be of the order of the local sound speed.}
{ Wind velocity constraints the relative strengths of wind mass and angular momentum loss and since it compares favorably with observations ($\sim$ a few tens of kilometers per second), it provides an additional sanity check for the disk wind model.}

\section{Summary of observational data}

\begin{table}[h]
\caption{Sources used for comparison with the models.}
\label{table:sfregions}
\begin{tabular}{llll}
\hline
Figure  &   Observed quantities        &   Star forming regions   &   References  \\ 
\hline 
Fig. \ref{fig:Rdiskevo} &  Gas radius    & Lupus       & \cite{Ansdell+18}                  \\
Fig. \ref{fig:Rdiskevo} &  Dust radius   & Ophiuchus, Taurus, Lupus, Chamaeleon  &  \cite{Hendler+20}, \cite{Manara+23}  \\
Fig. \ref{fig:Mdiskevo} &  Dust mass     & Ophiuchus, Taurus, Lupus, Chamaeleon  &   \cite{Manara+23}, \cite{Miotello+23}     \\
Fig. \ref{fig:Mdiskevo} &  Flux $^{13}$CO \& Flux C$^{18}$O  & Lupus, Chamaeleon  & \cite{Ansdell+16}, \cite{Long+17}   \\
Fig. \ref{fig:observed} &  Flux Band 6 \& $\alpha_{\rm mm}$   & Lupus  & \cite{Ansdell+18}    \\
Fig. \ref{fig:observed} &  Dust \& gas radius     & Lupus  & \cite{Ansdell+18}      \\
Fig. \ref{fig:RM}       &  Disk mass \& radius     & Lupus  & \cite{Ansdell+16}, \cite{Miotello+23}       \\
\hline
\end{tabular}
\end{table}

\end{document}